\documentclass{rQUF2e}

\usepackage{subfigure}
\usepackage{graphicx}
\usepackage{placeins}
\usepackage{amsmath}

\theoremstyle{plain}
\newtheorem{theorem}{Theorem}[section]

\theoremstyle{definition}
\newtheorem{definition}[theorem]{Definition}

\theoremstyle{remark}
\newtheorem{remark}{Remark}

\begin{document}


\title{Heavy-tailed features and dependence in limit order book volume profiles in futures markets}

\author{K. L. Richards$^{(\ast,1,2)}$ \thanks{$^\ast$Corresponding author.
Email: kylieanne.richards@gmail.com} and G. W. Peters$^{(3,4,5)}$ and W. T. M. Dunsmuir$^{1}$\\
\affil{$1$ School of Mathematics and Statistics, University of NSW, UNSW Sydney, NSW, 2052, Australia\\
$2$ Boronia Capital Pty. Ltd., 12 Holtermann street, Crows Nest, NSW, 2065, Australia\\
$3$ Department of Statistical Science, University College London, London, UK\\
$4$ Oxford Mann Institute, Oxford University, Oxford, UK\\
$5$ Systemic Risk Center, London School of Economics, London, UK.
} \received{v1.1 released March 2015} }

\maketitle

\begin{abstract}
Extensive literature on the properties of the Limit Order Book (LOB) has emerged with the access to ultra-high frequency data from electronic exchanges. The study of fundamental statistical attributes in such data plays an increasingly important role in aspects of financial modeling. This research is of particular relevance to trading strategies and best execution practices to satisfy the increasing proliferation of regulation. 

Only a limited number of studies have focused primarily on stochastic features of the volume process in the LOB, with the majority of studies centred on the price process. This paper investigates fundamental stochastic attributes of the random structures of the volume profiles in each level of the LOB. In particular, we investigate the ability to capture core features of the volume processes at different levels of depth under three families of models: $\alpha$-stable, generalized Pareto distribution and generalized extreme value and find that there is statistical evidence that heavy-tailed sub-exponential volume profiles occur on the LOB bid and ask and on both intra-day and inter-day time scales. In futures exchanges, the heavy tail features are not asset class dependent and they occur on ultra or mid-range high frequency data. Of the distributions and estimation methods considered, the generalized Pareto distribution MLE provided the best fit for all assets. We demonstrate the impact of the appropriate modeling of the heavy tailed volume profiles on a commonly used liquidity measure, XLM. In addition, utilizing the generalized Pareto distribution to model LOB volume profiles allows one to avoid over-estimating the round trip cost of trading and also avoids erroneous estimations of volume leading to significant LOB imbalances in less liquid assets. We conclude that building blocks for any volume forecasting model should account for heavy tails, time varying parameters and long memory present in the data.


\begin{classcode}Please provide at least one JEL Classification code\end{classcode}

\end{abstract}

\section{Introduction}
By utilizing new technologies, high frequency trading firms are able to execute electronic transactions within milliseconds or even microseconds and therefore take advantage of small price discrepancies. Typically their strategies may make as little as one basis point per trade. The very short time-frames used to enter and liquidate positions means that incorrect distributional assumptions underlying the forecasting models on available liquidity in the Limit Order Book (LOB), can very quickly cannibalize any profits made for that trade. Moving the price by under-estimating the liquidity, or alternatively, failing to execute enough volume when there is a larger liquidity event will degrade the quality of the trading strategy. Thus, key to the performance of any high frequency trading strategy is the underlying distributional assumptions made about the volumes on the LOB. The study of volume processes in the LOB also provides direct information regarding liquidity structures intra-daily in the LOB on a given exchange.

In addition, understanding fundamental features of the volume process at different depths of the LOB intra-daily are crucial in ensuring the standards of electronic exchange regulations are met with best practice. There are a range of new regulations being implemented throughout Europe and the US to further regulate the processing, placement and clearing of trades in electronic exchanges. These regulations, which in Europe fall under the ``Lamfalussy Directives'' include the Prospectus Directive, the Market Abuse Directive, the Transparency Directive and the highly anticipated Markets in Financial Instruments Directive (MiFID) \citep{Commission2014}. Under the last directive MiFID, the aim is to develop harmonized regulations for investment services across the 31 member states of the European Economic Area.

MiFID covers almost all tradable financial products with the exception of certain foreign exchange trades. The component of MiFID we believe that will be better understood by the type of analysis we undertake in this paper pertains to the key aspect of this directive known generically as ``Best Execution'' practice. In particular it states in \citet[section 2]{CESR2006} that
\begin{center}
`\textit{
MiFID's best execution regime requires investment firms to take all reasonable steps to 
obtain the best possible result for their clients, taking into account price, costs, speed, likelihood of 
execution and settlement, size, nature or any other consideration relevant to order execution. 
CESR considers this requirement to be of a general and overarching nature. 
}' 
\end{center}

Furthermore, and as noted by the Committee of European Securities Regulators in their white paper guide, \citet{CESR2006}, 
\begin{center}
`\textit{
	MiFID's best execution requirements are an important component of these investor protection 
	standards as they are designed to promote both market efficiency generally and the best possible 
	execution results for investors individually. [....] MiFID's best execution regime is set out as follows in the Directives. Article 21 of Level 1 and Articles 44 and 46 of Level 2 set out the requirements for investment firms that provide the  service of executing orders on behalf of clients for MiFID financial instruments and, indirectly via  Article 45(7), for investment firms that provide the service of portfolio management, when  executing decisions to deal on behalf of client portfolios. 
	}' 
\end{center} 

Understanding liquidity in the LOB is a core component of best execution. To provide clients with best execution regime under this directive, it is crucial that firms have accurate models of liquidity to accurately estimate market impact cost and timing risk which underpins best execution. 

The majority of the literature concerning the statistical properties of high frequency data primarily focuses on returns series of trade data, or price processes in the LOB. The limited numbers of studies that do consider volumes, do so in the framework of data been aggregated at a fixed grid of price points (ticks) away from the best bid and ask and typically not assessing intra and inter-day volume features. This work extends from papers such as, \citet{Biais1995}, \citet{Challet2001}, \citet{Maslov2001}, \citet{Bouchaud2002}, \citet{Gu2008} and \citet{Chakrabort2010}, \citet{Gould2012} by broadening the scope of assets considered, exploring the features of the data and making assessment of a range of flexible statistical models to capture these features. Previous studies have primarily focused on the two parameter (shape, scale) light tailed gamma distribution family which we demonstrate is limited in its ability to accurately capture the skewness and kurtosis features in the LOB volume data. Considerations of tail properties of the volume profile have not been explored in previous studies on LOB volume data. In addition, previous studies only consider assets from a single exchange, whereas this study considers futures comprised of different asset classes across five different exchanges. The markets considered in this study are electronic in nature and the focus will be on futures markets. The time frame of estimation of both the inter and intra-day level is across one year, a much more extensive study in terms of the cross-section of the sample used to test these novel distributional features compared to previous published studies of which we are aware.

In this research we consider the Market Depth (Level II) LOB volume profile time series data for a range of futures market assets as detailed in Table ~\ref{Tab:FuturesAssets}. A key feature of this study is the vast quantity of data utilized. To put this into context, for a single random day in 2010 GOLD had 814,580 events by 10 levels of volume in the LOB data; 5YTN had 790,006 events by 10 levels of volume at different prices in the LOB data and SP500 had 3,717,465 rows by 10 levels of LOB data. This study considers 250 trading days in the year of 2010. Markets considered in this study are classified as order-driven markets which constitute the most common form of market. \citet{Gould2012} provide a good description of the alternative type of market place, a quote-driven market and provide a mathematical description of the more flexible order driven market which we consider in this study. The data is constructed by taking a sub-sample of data, whereby the last volume of the LOB at a specified time increment is recorded. The volume is defined by the random vector $X_{t,d}^{i,j} \in \mathbb{R}$, where $i$ represents the level that the order is placed on the order book, $i \in \{-5, \dots,-1,1, \dots, 5\}$, $j$ is the asset, $t$ is the intra-day time and $d$ is the trading day. 

\begin{table}[h]\footnotesize
\begin{center}
\begin{minipage}{115mm}
\tbl{Asset description used in the analysis and modeling. Market hours refer to the liquid market hours in local trading time of the exchange.}
{\begin{tabular}{@{}lccc}\toprule
Asset Name  &Acronym &Liquid Market Hours (local time) & Exchange \\
\hline
Interest rate derivatives\\
\hline
1. 5 Year T-Note  &  \textit{5YTN}&  7:30:00 to 14:00:00 & CBOT \\
2. Euro-BOBL &  \textit{BOBL}&       8:00:00 to 19:00:00 & EUREX\\
\hline
Equity derivatives \\
\hline
3. SIMEX Nikkei 225 &\textit{NIKKEI} &    8:00:00 to 14:00:00 & SGX\\
4. E-mini S\&P 500  &  \textit{SP500} & 08:30:00 to 15:00:00 & CME\\
\hline
Precious Metals\\
\hline
5. Gold & \textit{GOLD} & 06:30:00 to 13:30:00 & COMEX\\
5. Silver & \textit{SILVER} & 08:30:00 to 13:00:00 & COMEX\\
\end{tabular}}
\label{Tab:FuturesAssets}
\end{minipage}
\end{center}
\end{table}
 
This study is designed to create the building blocks for a flexible dynamical model of LOB volume profiles. The first aim of this paper is to study the statistical properties of volume profiles using market depth (Type II) order book data to five levels on the bid and ask, totaling 10 levels of volume. We make inference on common features of several LOB multivariate stochastic structures, with the focus on sub-exponential behavior in the tails of the marginal distributions of the volume profiles for each level of depth on the bid and ask. We fit a range of flexible statistical models to the LOB volume profiles to investigate these features on an intra-day as well as an inter-day time scale. These models include the following families: generalized extreme value distributions, the generalized Pareto distributions and univariate $\alpha$-stable distributions. These parametric models developed for heavy tails in the continuous case have a well understood statistical interpretation and this will directly inform the statistical attributes of these stochastic volume processes on the LOB. To ensure the accuracy of the statistical and financial conclusions drawn from the analysis we consider several parameter estimation approaches for each model which include: generalized method of moment based approaches; empirical percentile based approaches; mixed maximum-likelihood and moment based methods; as well as L-moment based estimators. In the discussion of each method we comment on the suitability for practical estimation of such models using ultra-high frequency LOB data sets.

We also analyze a range of high frequency sampling rates, with sub-sampling frequencies of 10 seconds, 5 seconds, 2 seconds and 1 second intra-day for each trading day of 2010, however for compactness results are only presented for the 10 second sampling rate. These trading intervals provide a cross-section of frequencies that may be utilized by a high frequency trading strategy. By considering such sampling rates, we have two main objectives with the first being to disambiguate the real stochastic behavior of the heavy tailed features of the LOB volume profile structures from the high-frequency micro-structure noise, see recent work in \citet{ThesisMicroStruct}. The second is to understand, for a given market and asset class, whether basic statistical features such as heavy tailed attributes are persistent in the stochastic processes and possibly arising at different sampling rates. 

This research addresses long memory and autocorrelation for lower sampling rates and the impact of the trade-off between the variance in the parameter estimates due to reduced sample sizes and the bias introduced at the higher sampling rates. However, it is important to note that it is not the intention of this paper to specify the dependence structure.  

Finally, we investigate how our findings that demonstrate the importance of heavy tailed volume profile models, when compared to inappropriate lighter tailed distributional assumptions and focusing on the impact the modeling of liquidity, has on the LOB and the potential implications to a high frequency trading strategy. We conclude by making recommendations for key features that should be incorporated into a volume forecasting model. 

The remainder of the paper is organized as follows. Section two provides the motivation and evidence for heavy tails in the LOB volumes. This section starts by detailing the descriptive statistics of the six assets considered, the shape of the LOB volume profiles and long range dependence attributes. We then present results of the QQ-plots and mean excess plots which motivate the study of heavy tailed behavior. Section three gives a brief summary of the three models considered, $\alpha$-stable, generalized extreme value distribution and generalized Pareto distribution. This section provides a brief description of the different parameter estimation techniques used, with further detail given to the less well known robust approaches. Section four provides the empirical results for the three models and their different parameter estimation techniques applied to the LOB volume data and the effects of varying the sampling rates. The implications for dynamic modeling of LOB volumes is detailed. Section five presents a case study demonstrating the impact of considering heavy tailed features on a high frequency trading strategy. Section six summarizes the paper.

\section{Motivation and evidence for heavy tails on LOB volumes}\label{sec:Results1}

\subsection{The futures market data}

We consider limit orders only during \textit{liquid market hours} (Table ~\ref{Tab:FuturesAssets}), excluding auction periods and excluding days when the exchange is open on a public holiday. `Price discovery', which is the purpose of the continuous double-sided auction seen in the marketplace, becomes less efficient when liquidity is low \citet{Chordia2008}. This is driven by reported prices in the marketplace (at the best bid and ask) having insufficient volume during low liquidity times to support meaningful transaction sizes. Low liquidity indicates that few participants are currently active in the marketplace, hence the rate of information arrival is slow and the reported price becomes `inefficient'. The liquid market hours become those times of the trading day when the transaction rate in the marketplace is sufficiently high to guarantee that the price discovery process is operating `efficiently' and that the prices reported can be transacted in reasonable size.  

Liquid market hours in this context become those hours where there is sufficient volume through the market per unit time such that trades can be executed while still controlling for the cost of execution. As a general rule, the consumption of liquidity (ie, the portion of the market volume flow that is consumed in filling trades) should never exceed more than about $10\%$-$15\%$ of the volume available per unit time. The use of liquid market hours ultimately focuses the modeling of the volume dynamics on the stochastic volume profile attributes, rather than the short periods related to exchange specific pre-market open auction mechanisms. 

Table ~\ref{Tab:Desc} presents some descriptive statistics of the volumes at level one of the LOB for 2010 for each asset. We observe that the mean volume, the variance and the total daily spread of volumes on level 1 of the LOB for 5YTN, BOBL, SP500, NIKKEI, GOLD and SILVER can be large. All assets show high levels of positive skewness, with GOLD demonstrating the highest level (mean $\pm$ standard error) $7.61 \pm 4.88$ on the bid side and $8.74 \pm 7.04$ on the ask side. The mean level of kurtosis is high for all assets, however the standard deviation of kurtosis may indicate that high kurtosis is not always present in the data. For example, NIKKEI kurtosis is $6.5 \pm 3.9$ on the bid side and similarly for the ask side, $6.5 \pm 3.4$. The volumes for all assets on level one of the LOB start at one, $X_{t,d}^{ij} >0$.

\begin{table}[h]\footnotesize
\begin{center}
\begin{minipage}{115mm}
\tbl{Descriptive statistics for all trading days for 2010, using sub-sample data for level one of the LOB.}
{\begin{tabular}{@{}lcccccccc}\toprule
Asset & Side &    Max &   Min &     Median & Mean & Std &  Kurtosis &   Skew\\
\hline
5YTN &     Bid &   2553.05 &    1.05 &   390.88 &   420.19 &   324.95 &   8.73 &   1.30\\
     &     Ask &   2514.93 &   1.04 &    399.45 &   424.80 &   322.76 &   7.58&   1.20 \\
BOBL &     Bid &   2633.98 &   1.04 &   430.67 &   465.35 &   298.44 &   14.42&   1.59 \\
     &     Ask &   2606.95 &   1.04 &    429.43 &   462.18 &   293.88 &   14.42&   1.55  \\
SP500&     Bid &   4040.79 &   5.01 &   540.34 &   635.55 &   436.54 &   13.30 &   2.00 \\
     &     Ask &   4448.10 &   4.69 &   541.34 &   650.44 &   474.15 &   16.13 &   2.29 \\
NIKKEI&     Bid &   531.62 &   1.17 &   103.65 &   115.43 &   75.35 &   6.55    &   1.33\\
     &     Ask&   536.11   &   1.20 &   105.14 &   116.20 &   75.83 &   6.54    &   1.32\\
GOLD&     Bid &   172.96   &   1     &   4.62    &   6.78    &   8.66   &   139.65   &   7.61\\
     &    Ask&    206.85   &   1     &   4.62    &   6.80    &   9.64   &   184.71   &   8.74\\
SILVER&   Bid&    79.17    &   1     &   6.56    &   7.83    &   6.39   &   46.77    &   3.63\\
     &    Ask&    76.02    &   1     &   6.56    &   7.81    &   6.30   &   38.63    &   3.38\\
\end{tabular}}
\label{Tab:Desc}
\end{minipage}
\end{center}
\end{table}

\subsection{Shape of LOB volume profiles}
To assess the shape of the LOB volume profile, we develop a graphical representation of the volume on the LOB, which highlights particular empirical features that should be considered when developing statistical models for such stochastic structures. We construct a visualization that we denote as the volume profile for each asset obtained, by taking the median of the 10 second volumes for each hourly time increment throughout each trading day of the year 2010. In addition, the median volume per level on the LOB, levels 1 to 5, per day across the year of trading is considered. With this information we develop an understanding of the general volume features of the LOB, inclusive of depth considerations. 

\begin{figure}[h]
\begin{center}
\begin{minipage}{145mm}
{\subfigure[5YTN.]{
\resizebox*{7.5cm}{!}{\includegraphics{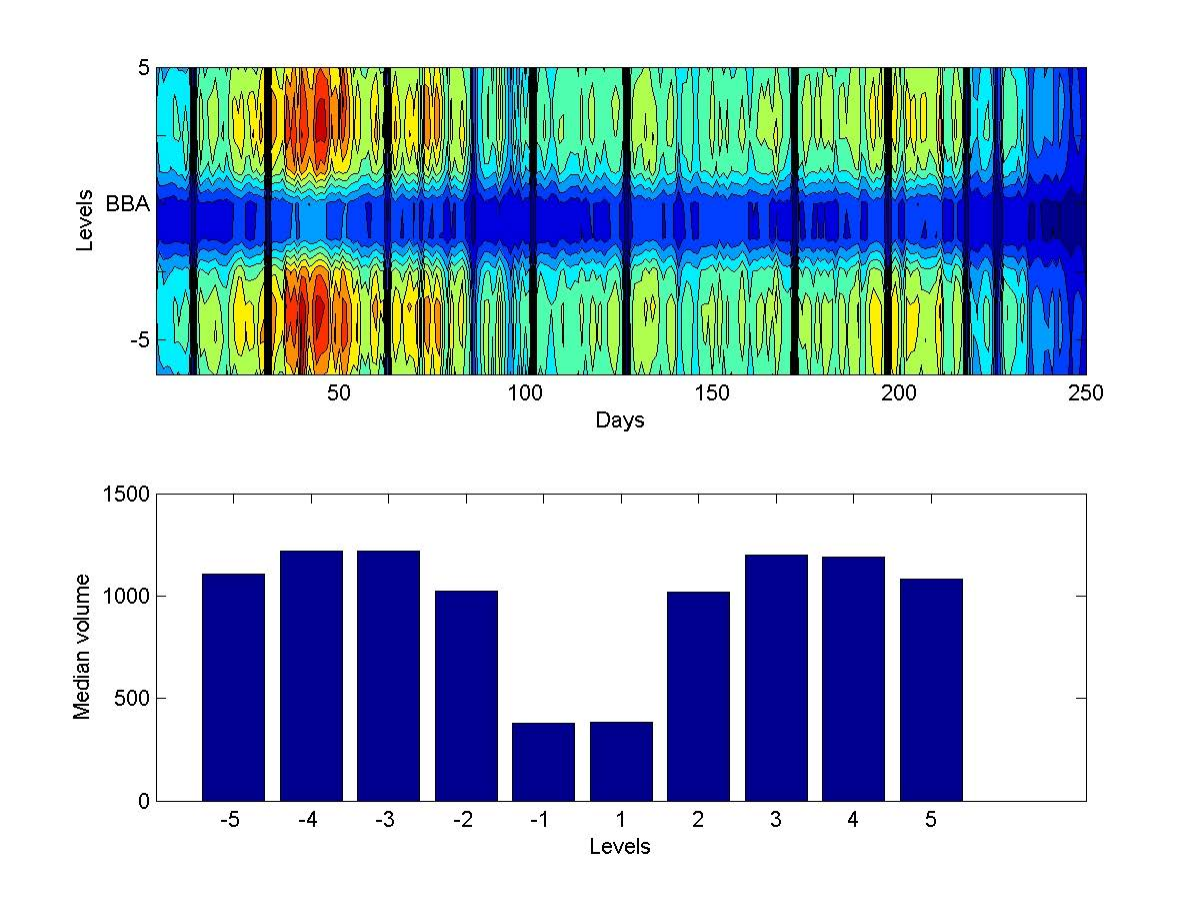}}\label{sample-figure_part_a}}
\subfigure[BOBL.]{
\resizebox*{7.5cm}{!}{\includegraphics{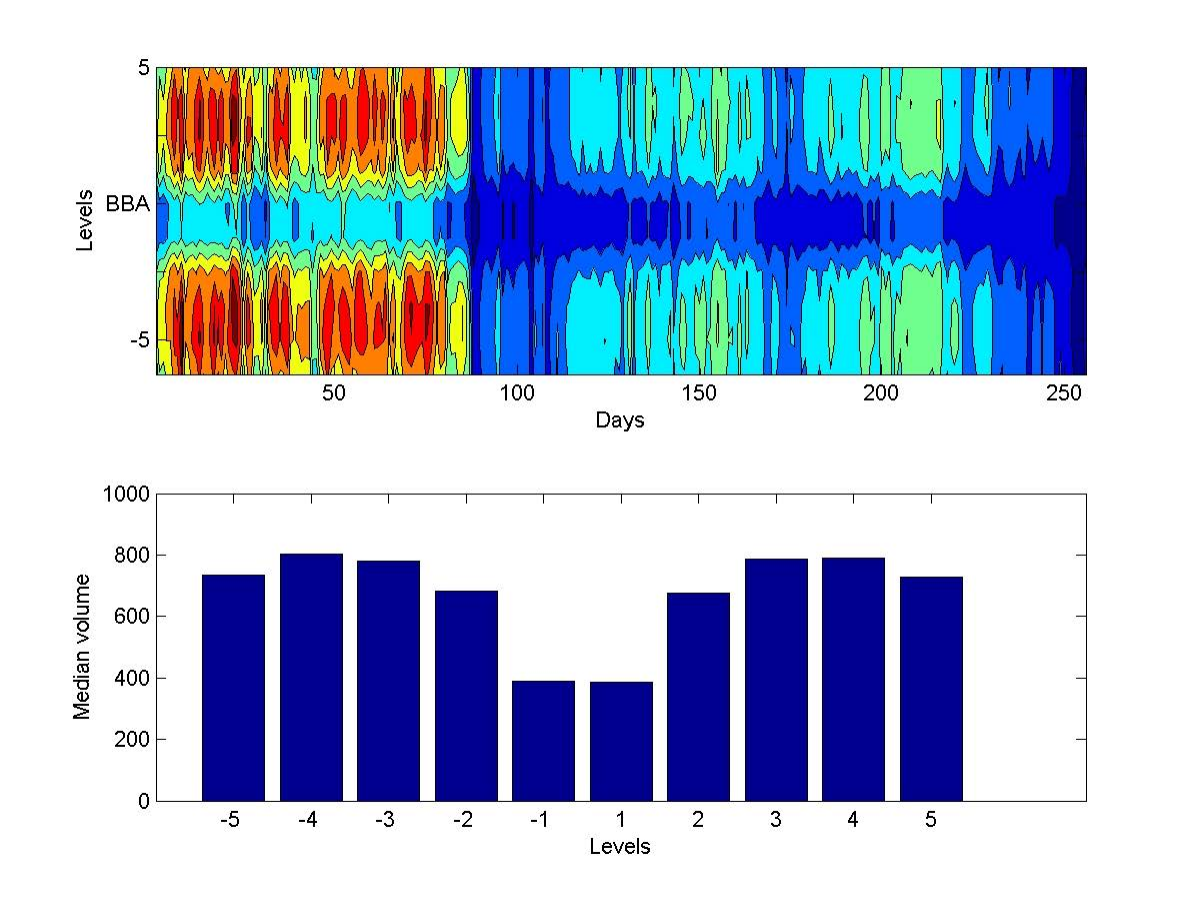}}\label{sample-figure_part_b}}
\caption{\label{fig2} Heat maps of the volume for the first five levels of the LOB and the median volume on each level of the LOB on the  bid and ask for 2010.}
\label{fig:BOBLVol}}
\end{minipage}
\end{center}
\end{figure}

The originally proposed idea of a LOB shape suggested a monotonically decreasing function away from the best bid and ask \citep{Biais1995,Bouchaud2002,Challet2001}. More recent findings from \citet{Potters2003}, \citet{Gu2008} and \citet{Chakrabort2010} suggested a humped shaped LOB. Consistent with the more recent findings, Figure \ref{fig:BOBLVol} represents a heat map of the intra-day volume profiles for the year and the common feature that appears between the 5YTN, BOBL, NIKKEI and SILVER is the humped shaped LOB. The SP500 and GOLD appears to have monotonically increasing volumes in the first 5 levels of the LOB. As shown in Figure ~\ref{fig:BOBLVol}, the heat chart for BOBL volumes are significantly higher at the start of the year and drop off towards the end of 2010. This feature is also present in NIKKEI and to a lesser extent, the 5YTN and SILVER. The SP500 and GOLD volumes tend to be relatively consistent throughout the year, contrary to the clear change in volume profile dynamic throughout the year of 2010 for several of the other key futures assets. 

The common feature that we observe in all assets is the inherent symmetry in the median volumes on the bid and ask at each level of the LOB. While the precious metals, GOLD and SILVER have significantly lower volumes placed on the LOB compared with other assets, they still demonstrate the inherent symmetry observed in the other assets.

\FloatBarrier

\subsection{Assessing long range dependence}

Long range dependence of volume profiles on the LOB is an important consideration in trading and MiFID's best execution regime when considering the likelihood of execution. If we observe persistence in the LOB volume profiles, this would be indicative of less trading activity or put another way, lower turnover of orders. 

Two methods are considered when studying the long range dependence of an asset, with the first being the well known Hurst exponent. We  then confirm our findings from the Hurst exponent analysis by also studying a more recent technique called the Extremogram, see \citet{davis2009}. 

\subsubsection{Hurst exponent (long memory) for the LOB volume profiles}
To study empirically the possibility of long memory in the LOB volume profiles at each of the 5 levels of volume on the bid and the ask, we first considered the autocorrelation function which suggested the presence of long memory for all assets. \citet{Gu2008}, utilized the Hurst exponent to test for long memory in the volume of the LOB by implementing a detrended fluctuation analysis to estimate the Hurst exponent on the 1 minute averaged volumes at the first 3 tick levels of the LOB. Detrended fluctuation analysis is a well-established scaling method for the detection of long-range correlations in time series \citep{Kantelhardt2002, Hu2001}. This provides an \textit{index of long-range dependence}, giving a quantitative measure of the relative tendency of a time series to regress strongly to the mean or cluster. As a guide, values for the index in the range $0.5 < H < 1$ indicate a time series with long-term positive autocorrelation \citep{Simonsen1998}. This indicates momentum in the intra-day volume profile, whereby high volume in the series is likely to be succeeded by another high volume period. Values of the Hurst exponent between $0 < H < 0.5$ indicates a time series with long-term switching between high and low volumes in adjacent 10 second time increments. 

\begin{figure}[h]
\begin{center}
\begin{minipage}{155mm}
{\includegraphics[height = 8.5cm, width = \textwidth]{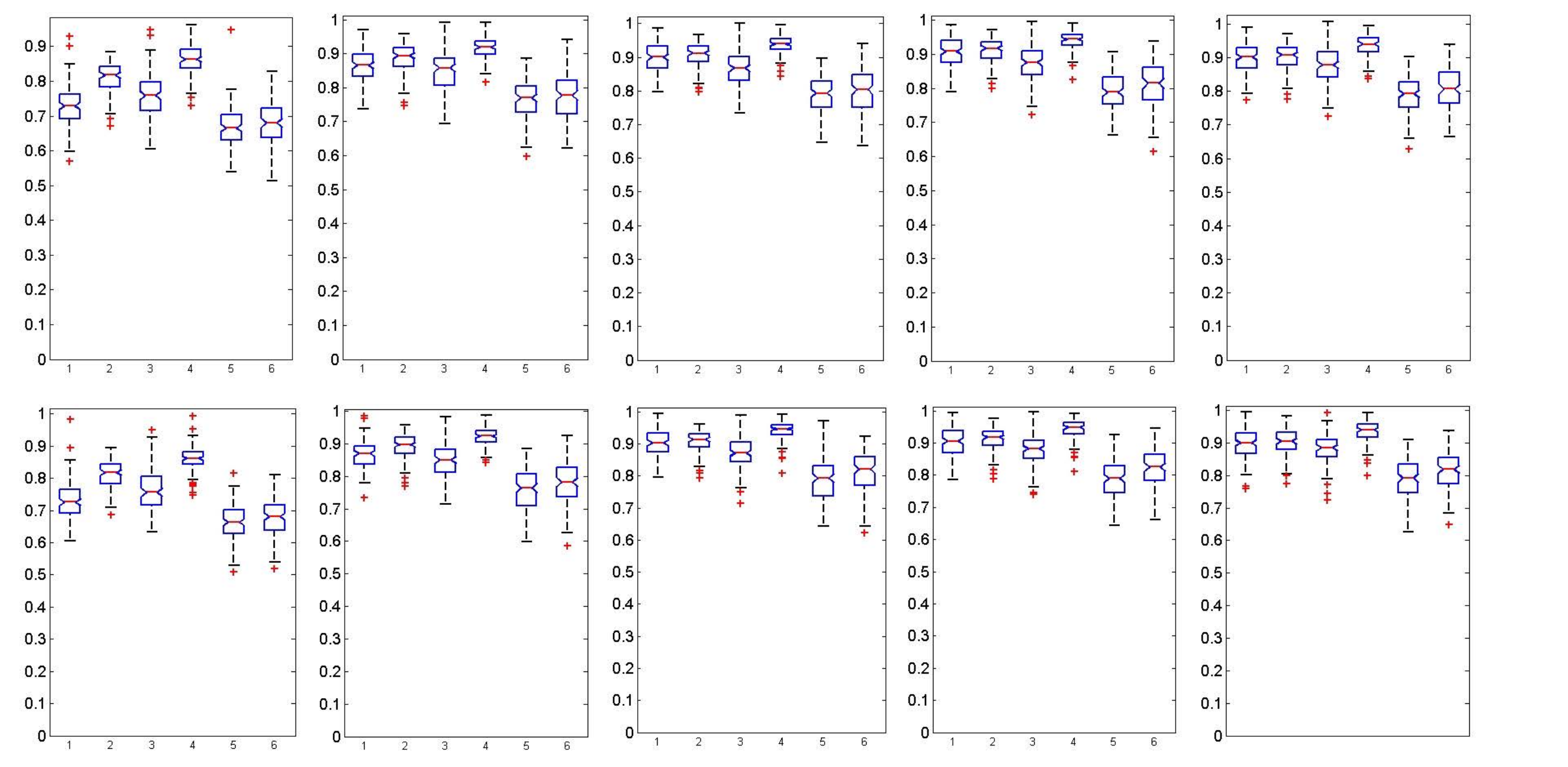}
  \caption{Boxplots of daily hurst exponent for 2010. \textbf{Top Row - Left to Right:} Level 1 to level 5 bid volumes; \textbf{Bottom Row - Left to Right:} Level 1 to level 5 ask volumes; \textbf{Each Subplot:} Assets left to right (1 to 6) are: 5YTN, BOBL, SP500, NIKKEI, GOLD, SILVER}
\label{fig:hurst}}
\end{minipage}
\end{center}
\end{figure}

With our aim to provide a richer data analysis, we implement the Hurst exponent estimation across a wider range of assets for every trading day of 2010, considering all 5 levels of the bid and ask. Figure \ref{fig:hurst} shows the Hurst exponents for each intra-day volume profile at level 1 to level 5 of the bid and ask in a sequence of box plots comprised of estimates for each trading day of 2010. In the analysis performed for each of the assets, we observed a range of results between $(0.65, 1)$ across all 5 levels on the bid and ask side volumes for all market sectors, exchanges and assets which is strongly consistent with what we would expect from data exhibiting long memory. Such features in general financial time series have previously been noted by \citet{lobato2000long} where they show that trading volume and volatility show the same type of long memory behavior. We have extended the range of assets and depth in the order book to show this long memory feature in the volume exists in a range of different asset classes in the futures market. If we relate this back to a high frequency trading strategy and fat tailed behavior, it is apparent that high liquidity events are likely to be followed by further high liquidity events, demonstrating potential momentum in the volumes on the order book. This forms a key consideration when modeling the dependence structure within dynamic models for the LOB volume profiles. 

\FloatBarrier

For each asset and each trading day, we also tested the volumes on each level for stationarity and in all cases the data was stationary at level 1 of the LOB for time increments of 10 seconds. When considering deeper levels of the LOB, level 2 and for assets SP500 and NIKKEI, we observed a stochastic trend, which was removed by first differencing, in approximately $1\%$ of trading days. Moving to level 3 of the LOB, the following assets exhibited non-stationarity with the indicated portion of days with this feature reported in brackets: 5YTN (bid $5.7\%$; ask $6.1\%$); BOBL (bid $0.39\%$; ask $0\%$); SP500 (bid $4.3\%$; ask $2.8\%$); NIKKEI (bid $6.9\%$; ask $6.5\%$). Level 4 demonstrates increasing days with a stochastic trend: 5YTN ($15.9\%$; $14.7\%$); BOBL (bid $0.39\%$; ask $0\%$); SP500 (bid $5.12\%$; ask $5.98\%$); NIKKEI (bid $8.57\%$ ask $8.88\%$). Level 5 of the LOB shows: 5YTN ($10.49\%$ bid $15.51\%$); SP500 (bid $7.17\%$ ask $2.08\%$); NIKKEI (bid $10.61\%$ bid $8.16\%$). As we investigate deeper levels of the LOB, we see that a stochastic trend is present and the long memory appears to increase.  However, the results for which a subset of the series required differencing due to stochastic trend removal, did not vary long memory findings significantly from the original Hurst exponent results presented in Figure \ref{fig:hurst}.

To further validate the results of this test, a randomized experiment for the Hurst exponent was implemented. For all assets across the 250 trading days considered, we found the Hurst exponent estimates to be robust to heavy tailed data.

Finally we consider the impact of varying the sub-sampled time increments on long memory for level 1 of the LOB. We consider a range of time intervals, one second to one minute, that are realistic for trading activities. Table ~\ref{tab:HurstSamp2} shows that in all cases we can see an increase in long memory as the time increment becomes finer. NIKKEI demonstrates the highest degree of long memory for the one second time increment, with a Hurst Exponent of 0.9. What is clear from this study is that for any realistic trading time interval, long range dependence is a persistent feature of volume profiles in the LOB and cannot be removed by considering lower frequency data. 
 
\begin{table}[h]\footnotesize
\begin{center}
\begin{minipage}{115mm}
\tbl{Mean Hurst Exponent across all trading days, using varying sub-sampled data for \textit{level 1} of the LOB }
{\begin{tabular}{p{1.4cm}p{1cm} p{2.5cm} p{2.5cm} p{2.5cm} p{2.5cm}p{2.5cm} }
\textbf{Asset} & \textbf{Side} &    \textbf{1 seconds} &    \textbf{2 seconds} &    \textbf{5 seconds} & \textbf{10 seconds} & \textbf{1 minute} \\
\hline
\textbf{5YTN} &     Bid &  0.7948 & 0.7765 &  0.7553  &0.7265 & 0.6286\\
     &     Ask & 0.7918& 0.7728& 0.7559 & 0.7276  & 0.6234\\
\textbf{BOBL} &     Bid &  0.8585 &  0.8460 &  0.8271 &0.8088 & 0.7463\\
     &     Ask & 0.8606 & 0.8483 & 0.8286 & 0.8105 & 0.7447\\
\textbf{SP500}&     Bid &   0.8206&   0.8038 & 0.7811  &0.7545 & 0.6524\\
     &     Ask &  0.8236  &  0.8070 &  0.7857 &0.7587 &0.6489 \\
\textbf{NIKKEI}&     Bid & 0.9020   & 0.8922 & 0.8764 & 0.8605& 0.8046 \\
     &     Ask&  0.9004  & 0.8918  &  0.8753 &0.8589 & 0.8042 \\
\textbf{GOLD}&     Bid &  0.7483  & 0.7295 &  0.6968& 0.6666& 0.5833\\
     &    Ask&  0.7452    &   0.7252  &  0.6915 & 0.6596 & 0.5730\\
\textbf{SILVER}&   Bid& 0.7651     &  0.7448  & 0.7070  & 0.6763 & 0.5802 \\
     &    Ask&  0.7633    &  0.7430   & 0.7048 & 0.6763 & 0.5830 \\
\end{tabular}}
\label{tab:HurstSamp2}
\end{minipage}
\end{center}
\end{table}

\subsubsection{Extremogram (serial dependence in the upper tail)}
In this section we apply a more recent technique called the extremogram, which was developed by \citet{davis2009} and provides a quantitative measure of dependence of extreme events in a stationary time series. Stationarity of the series was assessed in the previous section, with all LOB volumes exhibiting stationarity on level 1 of the LOB. 

For a strictly stationary $\mathbb{R}^d$ valued time series $(X_t)$, the extremogram is defined by 
\begin{equation}\label{eq:extre}
	\rho_{A,B}(h)=\lim_{x \rightarrow \infty} \mathbb{P}(x^{-1}X_h \in B| x^{-1}X_0 \in A), \text{ } h =0,1,2,\dots
\end{equation}

provided the limit exists. Because the volumes are positive, the \textit{extremogram} has been applied in this paper by choosing $A=B=[1,\infty)$. This reduces the extremogram to the upper tail dependence, which is often used in extreme value theory and quantitative risk management \citep{mcneil2005quantitative,Davis2012}.  

To estimate the extremogram, the limit on $x$ in ~\eqref{eq:extre} is replaced by a high quantile $(1-1/a_m)$ of the process \citep{Davis2012}. We select $a_m$ as the 20th percentile in order to be consistent with the peaks over threshold approach used when fitting the generalized Pareto distribution in later sections of the analysis. The sample extremogram, which is based on the observations $X_1,\dots,X_n$, is given by 
\begin{equation}\label{eq:samplext}
	\hat{\rho}_{A,B}(h)=\frac{\sum^{n-h}_{t=1}I(X_{t+h} \geq a_m, X_t \geq a_m)}{\sum^n_{t=1}I(X_t \geq a_m)}.
\end{equation}

\begin{figure}[h]
\begin{center}
\begin{minipage}{155mm}
{\subfigure[Bid side.]{
\resizebox*{8.5cm}{!}{\includegraphics{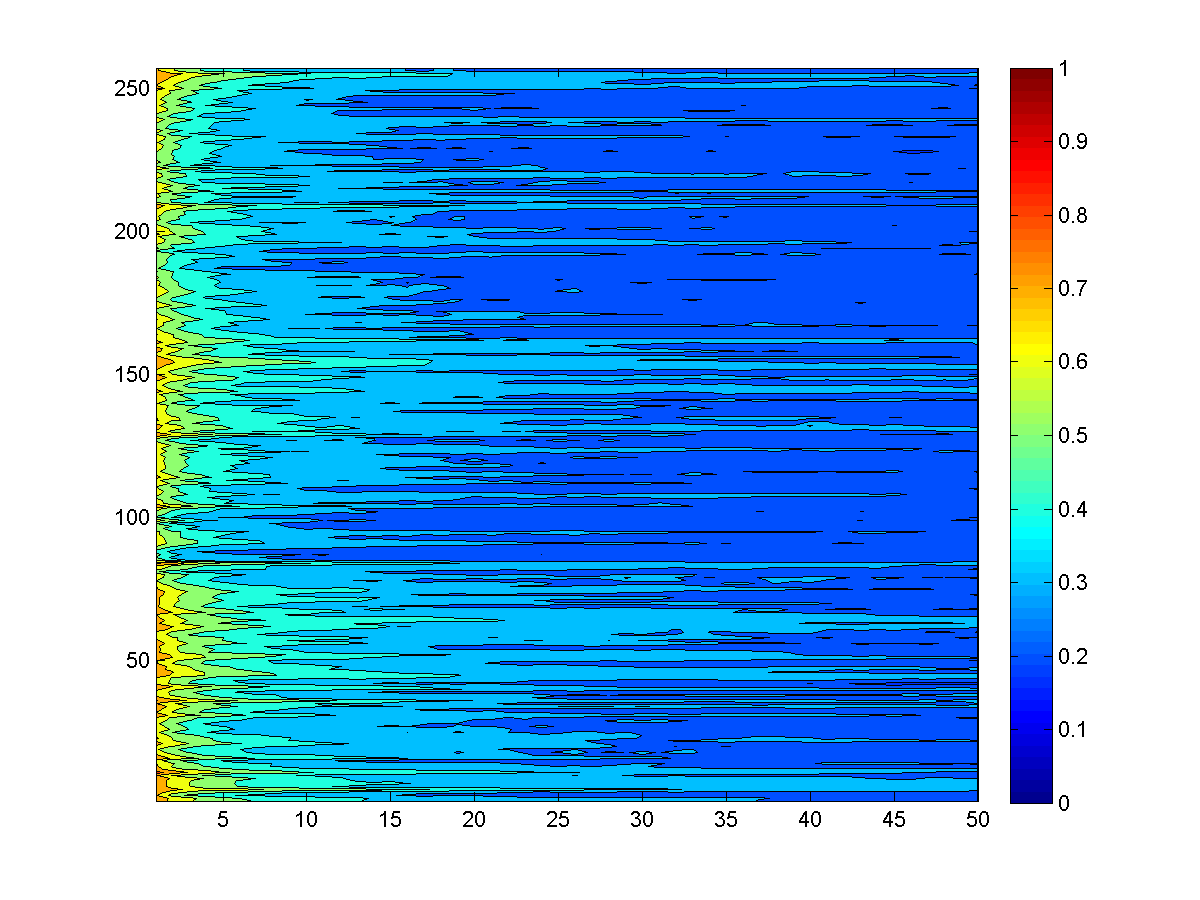}}\label{sample-figure_part_a}}
\subfigure[Ask side.]{
\resizebox*{8.5cm}{!}{\includegraphics{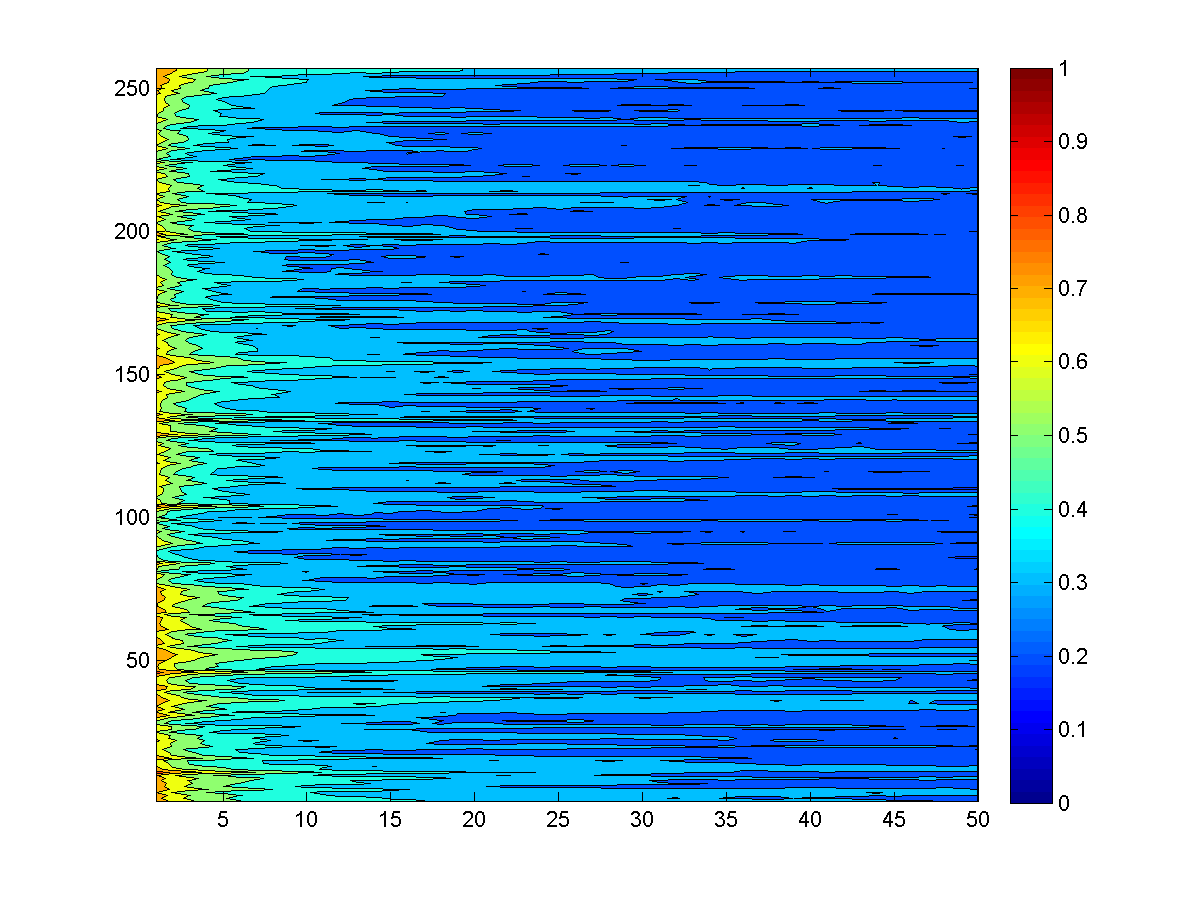}}\label{sample-figure_part_b}}
\caption{BOBL Extremogram heat map for 250 trading days using \textbf{10 second} sub-sampled data, $a_m = 80$th percentile, geometric distribution p-value$=1/200$.}
\label{fig:BOBLHeat}}
\end{minipage}
\end{center}
\end{figure}

The extremogram, being the conditional measure of extremal serial dependence is suitable for studying the persistence of a shock in the volume at a future time instant. The persistence of increased liquidity will allow a trader to increase their position in order to take advantage of the extra liquidity without incurring additional transaction costs associated with liquidity. BOBL (Figure ~\ref{fig:BOBLHeat}) and SP500 show persistent extremograms across all lags for the 250 trading days. SP500 and the 5YTN demonstrate lower levels of serial dependence compared with SP500. GOLD and SILVER show almost no serial dependence in the upper tails. 

\FloatBarrier

\subsubsection{The bootstrapped sample extremogram}
We consider the bootstrapped sample extremogram to measure the significance of the extremogram estimates. The method implemented is detailed by \citet{Davis2012}, who recommended the block re-sampling scheme which was introduced by \citet{Politis1994}. 

This method defines blocks by first choosing a starting point at random from a range $(1, n)$. The length of the block is choosen geometrically with a probability of $1/200$, as recommended by \citet{Davis2012}. The second and subsequent blocks are generated until the total length of the concatenated blocks is equal to or greater than our original sample size. The $95\%$ confidence intervals are produced by using $10,000$ stationary bootstrapped replicates and evaluating the indicator functions defined in \eqref{eq:samplext}. The bounds are found using the $0.025$ and $0.975$ quantiles from the empirical distribution of the bootstrapped replicates. This provides consistent estimators of the variability of the extremogram. 

When generating the confidence intervals for each of the assets across the 250 trading days, we consider a random selection of days for the bid and ask side. The solid horizontal line of height 0.20, represents the extremogram under an independence assumption. It is worth noting that \citet{Davis2012} use 0.04, however they had a very long time series of tens of thousands of observations, markedly larger than the few thousand data points of the time series used in this study. In addition, the 0.20 remains consistent with the peaks over threshold approach used when fitting the generalized Pareto distribution in later sections. If the solid horizontal line is well outside these confidence bands, this will confirm the serial extremal dependence in the upper tail.

\begin{figure}[h]
\begin{center}
\begin{minipage}{155mm}
{\includegraphics[height = 8.5cm, width = \textwidth]{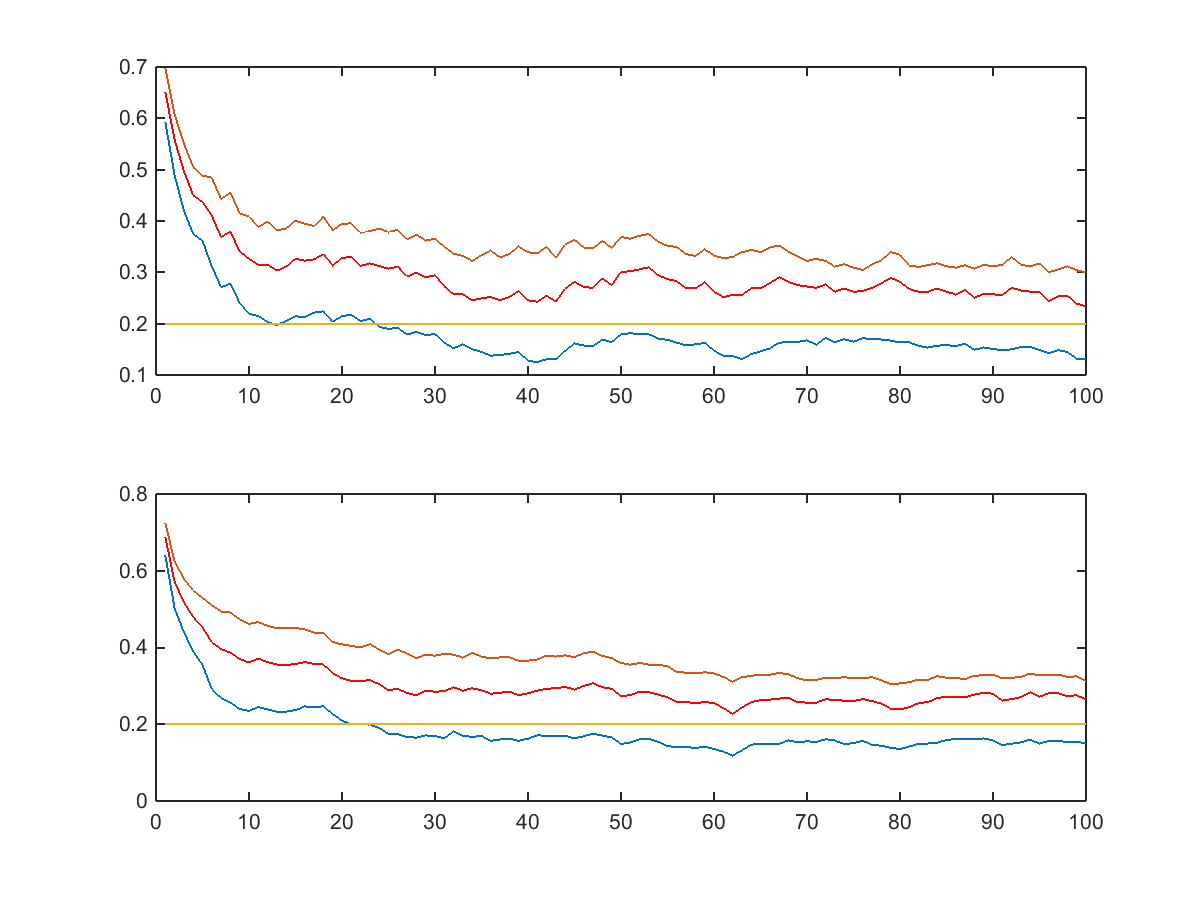}
\caption{Extremogram for BOBL using 10 second  sub-sampled data, 10,000 bootstrapped samples and for 100 lags. The upper charts show the bid-side and the lower charts show the ask-side.}
\label{fig:BOBLExtday12}}
\end{minipage}
\end{center}
\end{figure}

As demonstrated in Figure ~\ref{fig:BOBLExtday12}, BOBL has serial dependence in the upper tail up to the 25th lag for both bid and ask side at the $2.5\%$ level of significance. For the 5YTN and SP500 we observe serial dependence in the upper tail up to the 9th lag. The NIKKEI confirms serial dependence up to the 18th lag. Consistency of serial dependence across both bid and ask side exists for all assets. GOLD and SILVER shows serial upper tail dependence up to lag 3 only. It is worth highlighting that in all cases the median of the simulation always exceeds the independent and smoothly decays over time. This provides a strong indication of persistent long range dependence. The results for the Extremogram for each asset are largely consistent with what we have observed in the Hurst exponent, in that the GOLD and SILVER for the Hurst and Extremogram show the lowest levels of long memory and serial dependence in the upper tails, respectively. And conversely the NIKKEI and BOBL demonstrated the highest levels of long memory and serial dependence in the upper tails.   

Long range dependence has been noted by \citet{Gu2008}. By implementing two procedures, the Hurst exponent and the more recent technique being the extremogram, this research confirms this finding of long range dependence across all assets and with increasing dependence for shorter time increments. 

\FloatBarrier

\subsubsection{Extreme value theory and dependence}\label{sect:depend}
This section details the context in which one may apply elements of extreme value theory, typically applied to i.i.d. data sets, to a time series structure such as the volume processes on the bid and ask at each level of of the LOB. 

Volume spikes followed by additional volume spikes, being the presence of persistent heavy tailedness, is of particular importance to all algorithmic traders. Whether it be the agency broker seeking best execution, the market makers providing liquidity and the traders and arbitrageurs systematically increasing their positions to take advantage of the increased liquidity, they can all profit from a better understanding of the heavy tailed features of the volumes. When studying the heavy tailed features of the volume process, we adopt techniques from econometrics, statistics and probability, which includes working with appropriate aspects of  extreme value theory \citep{Beirlant2004}. Even with the time series structure observed in the volume process, it is appropriate and meaningful to consider extreme value theory. Considering both the marginal and the joint distributions of the process, as highlighted in the statistical finance paper of \citet{Cont2009}, is crucial in developing a holistic understanding of the volume process.  

For a generic volume process $\left(V_t\right)$ observed over time with the assumption that it is a strictly stationary time series, each individual observation of the volume process $V_t$ will have the same distribution function, denoted by $F_V$. If that data didn't have any time series structure we would be safe to assume that we could study, under the extreme value theory framework, the maximum of the series of i.i.d. random variables of length $n$, denoted $M_n = \max \{ Z_1, \ldots, Z_n\}$. One could then trivially show in this case that $\mathbb{P}\text{r}(M_n \le y) = \left\{\mathbb{P}\text{r}(V_j \le y) \right\}^n = F_V^n(y),$
where the independence of $V_t$ is used. 

For dependent data, such as data obtained from a time series structure, this relationship does not hold exactly and the distribution of the maximum $ M_n$ is not determined solely by the marginal $F_V$ alone, but rather from the complete distribution of the time series, i.e also the transition or conditional probabilities. However, it is also known in the extreme value theory literature that often a comparable, approximate extreme value theory relationship will be applicable, allowing for the application of extreme value theory in this context to time series data. In many settings the following approximate relationship will be applicable to aid in this context, $\mathbb{P}\text{r}(M_n \le y) \approx F_V^{n\eta}(y) \ge F_V^n(y)$, which will apply for large time series samples, as will be the case when considering ultra-high frequency LOB data sets intra-daily. For example, large $n$ with $\eta \in [0,1]$ denoting what will be termed the extremal index or extreme value index. The extremal index is a critical parameter in extreme value theory related to the heavy tailed nature of the data, see discussions in \citet{Embrechts1997}.

More precisely, in the independent case one can say that for $\tau \in [0,\infty]$ and every sequence of real numbers $\left(u_n\right)_{n\geq 1}$, then it holds that as $n \rightarrow \infty$ then $n \overline{F}_V(u_n) \to \tau$ iff $\mathbb{P}\text{r}(M_n \le u_n) \to e^ {-\tau}.$ From this statement one can say that the distribution $F_V$ of the volume process belongs to the domain of attraction of a generalized extreme value distribution.

In the context of a time series which does not display independence, one has approximately the following extension of this result. We consider the extreme value index $\eta \in [0,1]$, for the volume process time series $\left(V_n\right)$ when, for certain $\tau$ and sequence of real numbers $u_n$, one can show that $n\overline{F}_V(u_n) \to \tau$ and $\mathbb{P}\text{r}(M_n \le
u_n) \to e^ {-\eta \tau}.$ Furthermore, if an $\eta$ exists, then the value does not depend on the specific choice of $ \tau, u_n$. Therefore, using this approximate relationship between the distribution of the maximum and the exceedance probability, one obtains directly as $u_n$ grows large and 
$\overline{F}_V(u_n) \approx 0$, the following approximate relationship $\mathbb{P}\text{r}(M_n \le u_n) \approx e^{-\eta \tau} \approx e^{-\eta n \overline{F}_V(u_n)} \approx \left(1 - \overline{F}_V(u_n)\right)^{n\delta} =F^{n\delta}(u_n).$ In this context, one may adopt aspects of extreme value theory and apply it meaningfully to time series data. 

The data in this study demonstrates dependence, but this does not remove the necessity for one to consider the heavy tailed features present in the data, nor diminish the importance of a study on heavy tailed marginal distributions and ways in which these features are a crucial contribution to dynamical models of LOB volume profiles. The study of heavy tails in a marginal context is critical, as this has important effects on the accuracy of the estimators \citep{Francq2013}. Studies that consider heavy tailed features in financial market data include and are not limited to: \citet{Cont2001}; \citet{Francq2013}. In this paper we will use methods such as MLE, based on the assumption that observations are independent, hence the methods will not be efficient. However, we have large sample sizes on which to fit the extreme value distributions and it will be sufficient to obtain consistent estimators of the marginal distribution and the heavy tailed features. 

\subsection{Empirical evidence for heavy tails on LOB volumes}\label{sec:flexParaMods}

The literature on LOB modeling to date has considered only simple classes of two parameter shape-scale models for the statistical modeling of LOB data for price or volume. Papers by \citet{Bouchaud2002} and \citet{Gu2008} consider the distributional features of the LOB volumes and conclude that a two parameter shape-scale model given by a Gamma distribution is most suitable. However, from Table \ref{Tab:Desc} we observed high levels of positive skewness and kurtosis for all assets. These findings are indicative of heavy tails and since the volumes are strictly positive, we would expect heavy right tails. 

In this section we demonstrate the need for more sophisticated, flexible parametric models. We begin by fitting the gamma distribution as suggested in previous literature \citet{Bouchaud2002, Gu2008}. To obtain the estimators of the gamma distribution, we equated the population moments with the sample moments (moment matching). To assess the stability of the parameters and to assess how well the gamma distribution represents the skewness and kurtosis in the data, via moment matching, we estimated the parameters for all assets and for every time segment across an entire year of trading. From these parameters we estimated the mean, variance (which should be consistent with the sample estimates), skewness and kurtosis. For all assets, the gamma distribution provided a poor estimate of skewness and kurtosis.

The observations just made are likely to be due to the right tail of the volume distribution being heavier than can be obtained from the gamma distribution. We further investigate this particular aspect of the distribution by using two non-parametric techniques, with the first being the exponential quantile plot. This acts as a visual comparison between a \textit{medium sized tail} and allows us to identify a relatively \textit{fat-tailed} distribution. If the empirical CDF lie on the \textit{dashed} straight line then the volume profile is consistent intra-daily with an exponential distribution, however, the presence of a concave relationship, whereby the plot bends upwards away from a linear fit, indicates a fat-tailed distribution in the sub-exponential class.

\begin{figure}[h]
\begin{center}
\begin{minipage}{155mm}
{\includegraphics[height = 8.5cm, width = \textwidth]{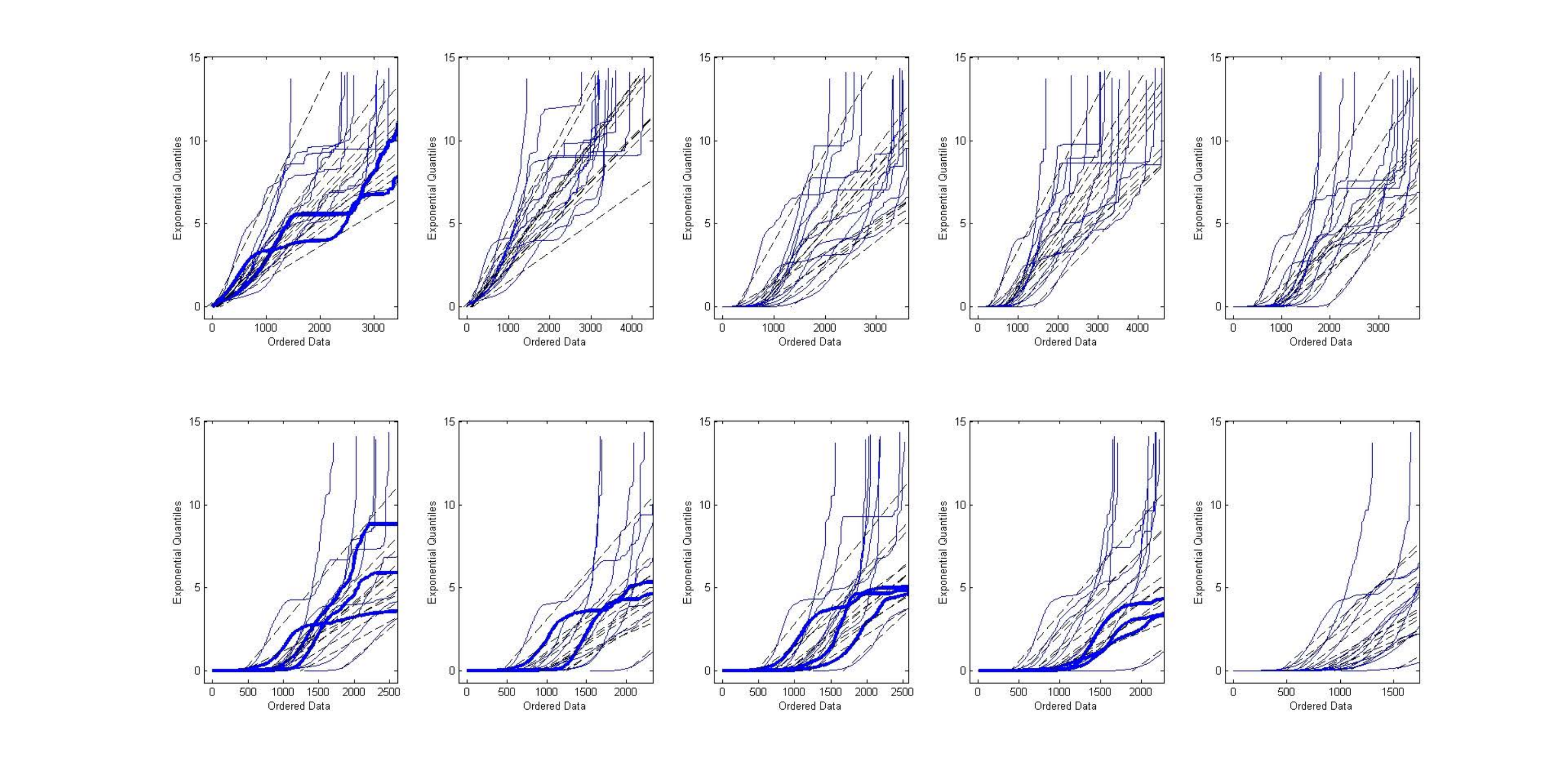}
  \caption{\textbf{5YTN:} Quantiles for exponential distribution model versus sample order statistics for intra-daily volume data every 25th trading day of 2010. \textbf{Top Row Bid} from left to right is Level 1 to Level 5 of LOB; \textbf{Bottom Row Ask} from left to right is Level 1 to Level 5 of LOB}
\label{fig:QQPlot5YrTNote}}
\end{minipage}
\end{center}
\end{figure}

Figure \ref{fig:QQPlot5YrTNote} shows the QQ-plot for the LOB data for the 5YTN relative to a generalized Pareto distribution with a tail index of $\gamma = 0$, making this a comparison between the right tail of the empirical CDF  and the right tailed exponential distribution.  The results presented are estimated intra-daily for every 25th trading day of the year and for each of the 5 levels of the LOB on the bid and ask sides. The choice of the 25th trading day provides an illustration of the general results we observed consistently for each trading day of the year, without overwhelming the visual representation and being an `extraordinary' trading day such as a futures expiry.  

The findings for the results in the QQ-plots comparing the empirical CDF  to the exponential model, demonstrate several interesting features. Firstly, starting with the 5YTN, in general at all levels of the volume profile there is a convex relationship relative to the exponential quantiles, indicating light tails for these profiles, see Figure \ref{fig:QQPlot5YrTNote}. However, on some days there is a clear evidence for a concave relationship between the empirical CDF and the quantiles of the exponential distribution, indicating the existence of a power law relationship. To help distinguish these days, we have emphasized examples of these particular trading days with a thicker solid line on the QQ plots. It is also clear from this analysis that there is a stronger tendency for power law relationships for the right tail of the volume profile on the ask side, relative to the bid side for the 5YTN. The BOBL also has occasional trading days which indicated the presence of heavy right tails for the intra-day volume profile indicative of a power law relationship. In this case it is clear that there is a stronger tendency for such heavy tailed features to occur on all 5 levels of the bid side throughout 2010, as opposed to the ask side which indicates far fewer examples of power law tails. The NIKKEI demonstrates occasional concave relationships for the right tail of the volume profile for example on level 1 of the ask and level 3 of the bid and ask. 

\begin{figure}[h]
\begin{center}
\begin{minipage}{155mm}
{\includegraphics[height = 8.5cm, width = \textwidth]{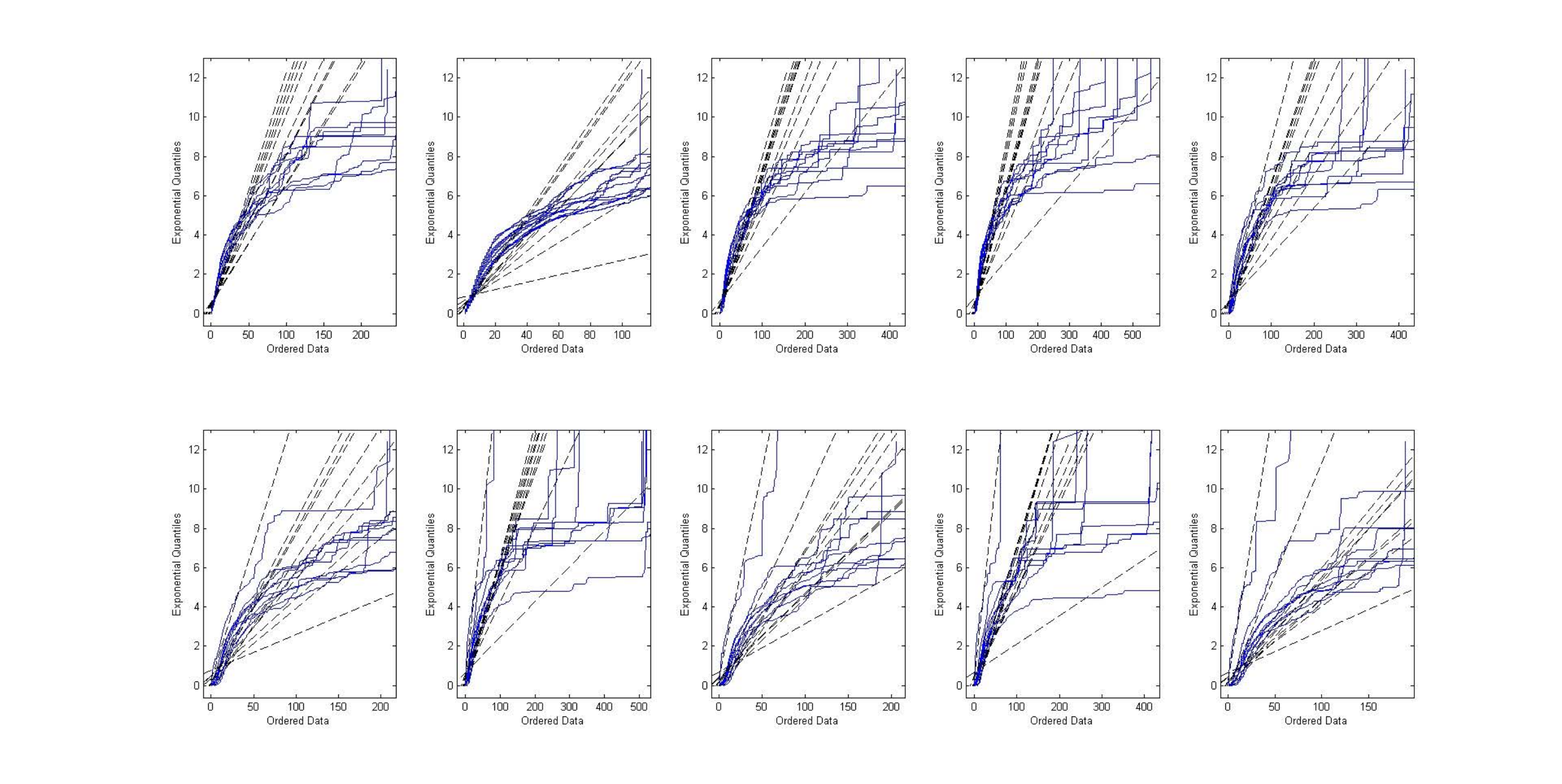}
  \caption{\textbf{GOLD:} Quantiles for exponential distribution model versus sample order statistics for intra-daily volume data every 25th trading day of 2010. \textbf{Top Row Bid} from left to right is Level 1 to Level 5 of LOB; \textbf{Bottom Row Ask} from left to right is Level 1 to Level 5 of LOB}
\label{fig:QQPlotGold}}
\end{minipage}
\end{center}
\end{figure}

For the SP500 it is apparent that the power law relationship in the tails is prominent more often in the volume profile at all levels 1 to 5 in both the bid and ask sides. The attributes of futures contracts on GOLD (Figure ~\ref{fig:QQPlotGold}) indicates strongly the presence of heavy tailed relationships on every trading day analyzed on all levels of the LOB for the bid and ask. SILVER similarly has consistent evidence of heavy tailed relationships in the right tail of the intra-day volume profile.

\FloatBarrier

\FloatBarrier

The second technique we consider, is the mean excess plot to aid the assessment of heavy tailed behavior, see \citet{kratz1996qq}. We proceed by presenting the mean excess plot intra-daily for every 25th trading day of the year and for each of the 5 levels of the LOB on the bid and ask sides at ten second sub-sampling frequency, for each asset. The sample mean excess function defined by Equation \eqref{Eqn:SampleMeanExcess}, represents the sum of the excesses over a threshold $u$ divided by the number of data points which exceed the threshold $u$. It approximates the mean excess function describing the expected exceedence amount for a particular threshold $u$ given an exceedence in the volume profile has occurred. If the empirical mean exceedence function estimate has a positive slope for large thresholds $u$ then this indicates that the observed volume profile data is consistent with a generalized Pareto distribution with a positive tail index parameter \citet[Chapter 1]{Beirlant2004}. Worth noting is that the mean excess function is sensitive to the larger thresholds when the corresponding $e_n(u)$ defined in Equation \eqref{Eqn:SampleMeanExcess} may contain fewer observations. \citet{Embrechts1997} suggest in this case to ignore the larger thresholds, which is not problematic when considering the large data sets of the LOB. 

The sample mean excess is given by,
\begin{equation} \label{Eqn:SampleMeanExcess}
e_n(u) = \frac{\sum_{i=1}^n (X_i - u)\mathbb{I}_{\left\{X_i > u\right\}}}{\sum_{i=1}^n \mathbb{I}_{\left\{X_i > u\right\}}},
\end{equation}
which estimates the conditional expectation $e(u) = \mathbb{E}\left[(X-u)| X>u\right]$.

\begin{figure}[h]
\begin{center}
\begin{minipage}{155mm}
{\includegraphics[height = 8.5cm, width = \textwidth]{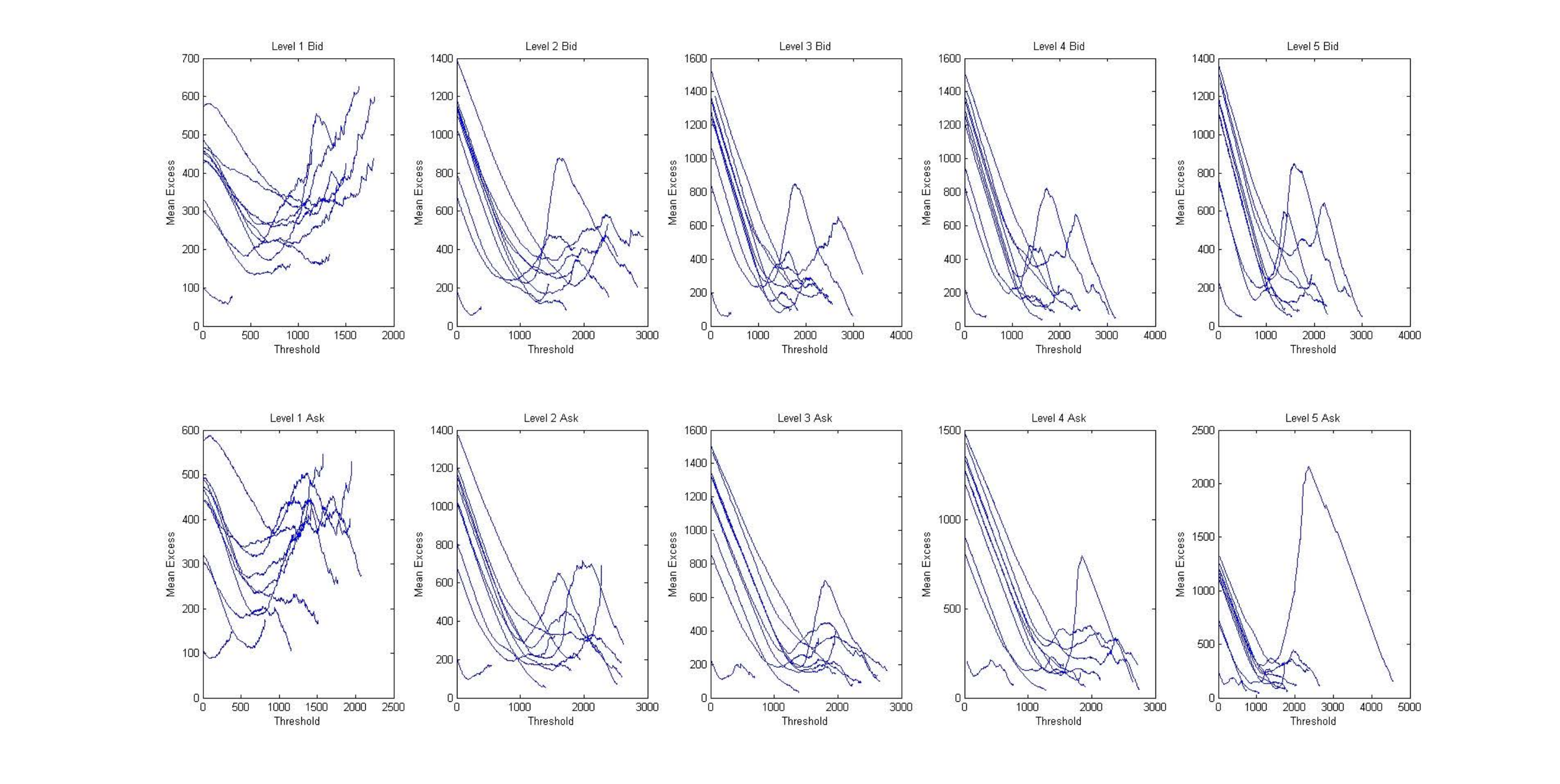}
  \caption{\textbf{5YTN:} Mean Excess plot verus the threshold $u$ for intra-daily volume data every 25th trading day of 2010. \textbf{Top Row Bid} from left to right is Level 1 to Level 5 of LOB; \textbf{Bottom Row Ask} from left to right is Level 1 to Level 5 of LOB}
\label{fig:MEPlot5YrTNote}}
\end{minipage}
\end{center}
\end{figure}

\begin{figure}[h]
\begin{center}
\begin{minipage}{155mm}
{\includegraphics[height = 8.5cm, width = \textwidth]{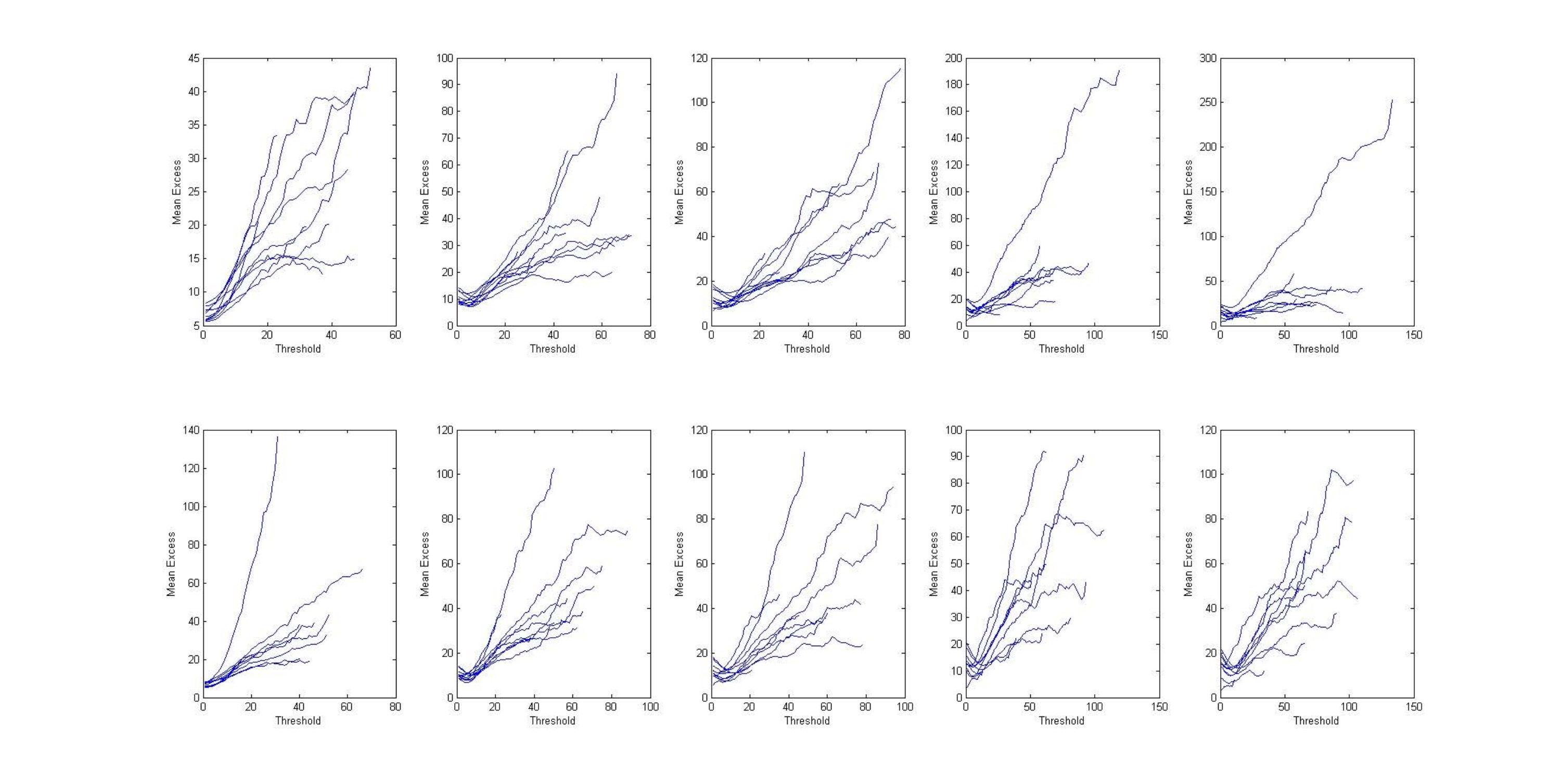}
  \caption{\textbf{GOLD:} Mean Excess plot verus the threshold $u$  for intra-daily volume data every 25th trading day of 2010. \textbf{Top Row Bid} from left to right is Level 1 to Level 5 of LOB; \textbf{Bottom Row Ask} from left to right is Level 1 to Level 5 of LOB}
\label{fig:MEPlotGold}}
\end{minipage}
\end{center}
\end{figure}

In Figure \ref{fig:MEPlot5YrTNote} we observe the Mean Excess plot versus the threshold, $u$, for the 5YTN. It indicates a clear upward trend as the threshold (x-axis) exceeds $500$ for all of the trading days explored on the bid at level one, consistently indicating the presence of heavy tailed power law relationships in the volume profile. At level 1 of the ask there is a mix between evidence for some days having heavy tailed attributes in the right tail of the volume profile and other trading days with lighter tails in the higher threshold region. This is also present throughout the other levels of the LOB on the bid and ask. The results for the BOBL and the SP500 demonstrate on several of the trading days, a strong indication of power law relationships in the right tail of the volume profile. SP500 is very pronounced at level 1 of both the bid and ask. The NIKKEI also indicates the occasional presence of power law right tail. As expected from the QQ plots, GOLD (Figure ~\ref{fig:MEPlotGold}) and SILVER indicate strong power law relationships consistently in the intra-day volume profiles on the majority of trading days presented.

\FloatBarrier

\FloatBarrier

In conclusion, we have convincingly demonstrated that the previously suggested gamma distribution is not capable of capturing the heavy tailed nature of the distribution of volumes in the LOB at all levels. Tails are heavier than exponential and more consistent with the power law. In the next section we assess alternative distributions that are better able to capture this heavy tailed behavior. 

\section{Distributional models, methods of estimation and testing}
The following section presents the parametric families of models that we consider for modeling levels 1 to 5 of the LOB volume profiles. Stable distributions have been proposed as a description for the large data sets of the LOB volume data, due to their flexible heavy tail behavior and asymetry relationships. We also consider a second sub group of the sub-exponential family of models being the extreme value distributions, which have a well established statistical theory and give a greater flexibility in capturing heavy tailed features. Both models have asymptotic power law tails.

\subsubsection{A brief summary of $\alpha$-stable models}
Models constructed with $\alpha$-stable distributions possess several useful properties, including infinite variance, skewness and heavy tails \citep{Zolotarev1986,  Alder1998,Samorodnitsky1994,Nolan2007}. Considered as generalizations of the Gaussian distribution, they are defined as the class of location-scale distributions which are closed under convolutions. $\alpha$-stable distributions have found application in many areas of statistics, finance and signal processing engineering as models for impulsive, heavy tailed noise processes \citep{Mandelbrot1960,Fama1965,Fama1968,Nikias1995,Godsill2000,Melchiori2006,Peters2010,Peters2011,peters2012generalized,gu2012receiver}. Here we consider using this class for the modeling of the volume profiles in a LOB stochastic process. 

The univariate $\alpha$-stable distribution is typically specified by four parameters \citep{Levy1924}. The tail index $\alpha \in (0, 2]$ determining the rate of tail decay, $\beta \in [-1, 1]$ determining the degree and sign of asymmetry (skewness), $\gamma > 0$ the scale (under some parameterizations) and $\delta \in \mathbb{R}$ the location. The parameter $\alpha$ is typically termed the characteristic exponent, with small and large $\alpha$ implying heavy and light tails respectively and $\alpha<2$ having infinite variance. Gaussian ($\alpha = 2, \beta = 0$) and Cauchy ($\alpha = 1, \beta = 0$) distributions provide the only simple analytically tractable sub members of this family. In general, as $\alpha$-stable models admit no closed form expression for the density which can be evaluated pointwise, inference typically proceeds via the characteristic function. 

For modelling the LOB data, we use the following parameterization in which a random variable $X$ is said to have a stable distribution if its characteristic function has the following form:
\begin{equation} \mathbb{E}[\text{exp}(i\theta X)] = 	\left\{              
			\begin{array}{ll}
				\text{exp}\{ -\gamma^\alpha|\theta|^\alpha(1+i\beta(\text{sign}(\theta))\tan({\pi \alpha \over 2})(|\gamma \theta|^{1-\alpha}-1))+i\delta \theta\} & \text{, if   } \alpha \neq 1,\\ 
				\text{exp}\{ -\gamma|\theta|(1+i\beta({2 \over \pi})(\text{sign}(\theta))\text{ln}(\gamma|\theta|))+i\delta \theta\} & \text{, if   } \alpha = 1.
			\end{array}
			\right.
	\end{equation}\label{eq:alphaCF}

The series expansions for the corresponding density and CDF are provided in \citet{zolotarev1983univariate}.

\subsubsection{A brief summary of extreme value models}\label{subsubsect:EVTvolProf}
We present the characterization of the extreme value theory families considered for modeling of the right tail of the volume distribution at each level of the LOB profile. We focus on the generalized extreme value and the generalized Pareto distribution families of distributions.

\begin{definition}[Generalized extreme value distribution]{\label{Defn:GEVDist} The generalized extreme value distribution is defined by 
\begin{equation}
\mathbb{P}\text{r}\left(X < x;\mu,\sigma,\gamma\right) =  \exp\left\{-\left[1+\gamma\left(\frac{x-\mu}{\sigma}\right)\right]^{-1/\gamma}\right\},
\end{equation}
for $1 + \gamma(x - \mu) / \sigma > 0$, where $\mu\in\mathbb R$ is the location parameter, $\sigma > 0$ the scale parameter and $\gamma\in\mathbb R$ the shape parameter.
Furthermore, the density function is given by,
\begin{equation}
f(x;\mu,\sigma,\gamma) = \frac{1}{\sigma}\left[1+\gamma\left(\frac{x-\mu}{\sigma}\right)\right]^{(-1/\gamma)-1} \exp\left\{-\left[1+\gamma\left(\frac{x-\mu}{\sigma}\right)\right]^{-1/\gamma}\right\}.
\end{equation}
In addition the support of a random variable $X \sim H_{\gamma}\left(\frac{x-\mu}{\sigma}\right)$ is given by
\begin{equation}
S_X = \begin{cases}
\left[\mu - \frac{\sigma}{\gamma},\infty\right], \; \gamma > 0, \\
\left[-\infty ,\infty\right], \; \gamma = 0, \\
\left[-\infty, \mu - \frac{\sigma}{\gamma}\right], \; \gamma < 0. 
\end{cases}
\end{equation}
}
\end{definition}

The estimation of the generalized extreme value model parameters involves a block maximum based analysis with its associated estimation procedures and properties, see \citet{Beirlant2004}, \citet{bensalah2000steps} and \citet{embrechts1999extreme}. In short, the block maximum approach is a data preparation procedure that involves taking the maximum volume recorded for that LOB level within the sub-sample time increment used. The preparation of the volume profile data for fitting the generalized extreme value model involved taking intra-daily data for each asset over the period 2010 and splitting the data into blocks. The maximum volume submitted in the sub-sample time block is recorded, producing a set of $K$ ordered realized observations $\left\{x_{(1,K)},\ldots,x_{(K,K)}\right\}$. Comparing this to the $\alpha$-stable preparation of the data with $K$ samples, we see the key difference is that we use the maximum volume rather than the last volume recorded in the specified time increment when constructing the observed time series data.  

An alternative specification of such extreme value theory models is the peaks over threshold based formulation which is used for the generalized Pareto distribution model. For details of alternative data preparation and parameter estimation procedures see \citet{Beirlant2004}. In the case of peaks over threshold, the last observation in the time increment is recorded. Furthermore, we then retain the data that sits above a prespecified threshold. If the threshold was $120$, for example, we would not include this observation in the model. 

\begin{definition}[Generalized Pareto Distribution]{\label{Defn:GPDDist} A random variable $X \sim GP\left(\gamma,\sigma\right)$ to have a distribution and density (conditional upon translation to the origin - location parameter $\mu = 0$) given by
\begin{equation}
\begin{split}
F_X(x;\gamma,\sigma) &= \mathbb{P}\mathrm{r}\left(X < x|X \geq \mu\right) = \begin{cases}
1 - \left(1 + \frac{\gamma x}{\sigma}\right)^{-\frac{1}{\gamma}}, & \gamma \neq 0, \\
1 - \exp\left(- \frac{x}{\sigma}\right), & \gamma = 0, \\
\end{cases}
\end{split}
\end{equation}
\begin{equation}
f_X(x;\gamma,\sigma) = \frac{\sigma^{\frac{1}{\gamma}}}{\left(\sigma + \gamma x\right)^{\frac{1}{\gamma}+1}},
\end{equation}
with shape parameter $\gamma \in \mathbb{R}$ and scale parameter $\sigma \in (0,\infty)$. In addition, the support of the density is given by
\begin{equation}
S_X = \begin{cases}
[\mu,\infty) , \; \gamma \geq 0, \\
\left[\mu, \mu - \frac{\sigma}{\gamma}\right], \; \gamma < 0.
\end{cases}
\end{equation}
}
\end{definition}
					
\subsection{Statistical estimation for models of LOB volume profiles}\label{sect:param est}

There are numerous approaches that can be applied when fitting heavy tailed and flexible families of models such as the generalized Pareto distribution, generalized extreme value and $\alpha$-stable. Each approach will have different merits related to statistical efficiency, bias and variance trade-offs and importantly for the setting of analysis of LOB data (massive data sets) the methods being computationally robust and efficient. Computational robustness refers to computer science based robustness, whereby the algorithm that implements the statistical techniques will need to continue to operate despite abnormalities in data inputs. Computationally efficient refers to the amount or resources used by the methods and the time it takes to obtain results. This is an important consideration for high frequency trading strategies, as they often trade many assets across multiple markets and robust estimation methods which can be achieved in the high frequency time frames, become crucial in mitigating the risk in trading.   

In the cases where we discuss the Maximum Likelihood Estimation (MLE), we base this on the assumption of independence and as noted in the Section \ref{sect:depend} (last paragraph), MLE won't be statistically efficient, but it will be consistent. Efficiency considerations are secondary to our main objective of determining consistent estimators of distribution parameters. 

In Table ~\ref{tab:dist} we provide a summary of the methods utilized for each of the distributions. This table shows the distributions and methods used for volume data on level one of the LOB for each of the six assets: 5YTN, BOBL, SP500, NIKKEI, GOLD and SILVER. For each asset we consider varying time resolutions of 1 second, 2 seconds, 5 seconds and 10 seconds across each trading day in 2010. The discussion that follows will provide insight into the dynamics of the parameters over time, varying time resolutions, different estimation methodologies and associated implementation issues. In particular we provide details of several less well known but efficient and statistically robust approaches.

\begin{table}[h]\footnotesize
\begin{center}
\begin{minipage}{115mm}
\tbl{Distributions and methods fitted to volume data for six assets: 5YTN, BOBL, SP500, NIKKEI, GOLD and SILVER. MLE; Method of moments; McCullochs quantile based estimation \cite{McCulloch1986}; Pickands' estimator \cite{pickands1975statistical}; Empirical percentile method}
				{\begin{tabular}{p{3cm}p{0.9cm}p{1.5cm}p{1.6cm}p{1.4cm}p{1.4cm}p{0.9cm}}
				&&& \textbf{Method} \\
				\hline
				\textbf{Distribution}  & MLE  & McCullochs & Mixed L-moments & Pickands' & Empirical percentile method \\
				\hline
				1. $\alpha$-stable &     &   $\surd$ & & & \\
				2. Generalized extreme value  &  $\surd$     &  & $\surd$ & & \\
				3. Generalized Pareto distribution  &  $\surd$    &  & & $\surd$  & $\surd$  \\
			\end{tabular}}
			\label{tab:dist}
\end{minipage}
\end{center}
\end{table}

\subsubsection{$\alpha$-stable distribution parameter estimation}\label{subsect:aSTAB}

The method used to estimate the parameters for the $\alpha$-stable is based on the quantile approach of \citet{McCulloch1986} and \citet{McCulloch1998}. This approach was selected because it is known to be computationally robust and efficient. Estimates of the model parameters are based on sample quantiles, while correcting for estimator bias due to the evaluation of the sample quantiles. We let $x_{(p)}$ be the $p$-th population quantile and $\widehat{x}_{(p)}$ to be the sample quantile from the order statistics of the sample. The four parameters in the $\alpha$-stable model under the parametrization presented are determined from a set of five pre-determined quantiles for the parameter ranges $\alpha \in [0,2.0]$, $\beta \in [-1,1]$, $\gamma \in [0,\infty)$ and $\delta \in \mathbb{R}$ as detailed in \citet{McCulloch1986}. The stages of this estimation involve the following details:

\begin{enumerate}[Step 1]
	\item{\textbf{Obtain a finite sample consistent estimator of quantiles:} with the $x_{i}$ arranged in ascending order, the skewness correction is made by matching the sample order statistics with $\widehat{x}_{s(i)}$ where $s(i) = \frac{2i-1}{2n}$. Then a linear interpolation to $p$ from the two adjacent $s(i)$ values is used to establish $\widehat{x}_{(p)}$ as a consistent estimator of the true quantiles. This corrects for spurious skewness present in finite samples and $\widehat{x}_{(p)}$ is a consistent estimator of $x_{(p)}$.}
	\item{\textbf{Obtain estimates of tail index $\alpha$ and skewness parameter $\beta$:} in \citet{McCulloch1986} non-linear functions of $\nu_\alpha$ of $\alpha$ and $\nu_\beta$ of $\beta$ are provided in terms of the quantiles as detailed in Equation \eqref{Eqn:McCulloch86}.
\begin{equation} \label{Eqn:McCulloch86}
\nu_{\alpha}=\frac{x_{(0.95)}- x_{(0.05)}}{x_{(0.75)}-x_{(0.25)}} ,\; \nu_{\beta}=\frac{x_{(0.95)}+x_{(0.05)}-2x_{(0.5)}}{x_{(0.95)} - x_{(0.05)}}.
\end{equation}
We can therefore estimate these quantities $\widehat{\nu}_{\alpha}$ and $\widehat{\nu}_{\beta}$ using the sample estimates of the quantiles $\widehat{x}_{(p)}$. Now to obtain the actual parameters estimates $\widehat{\alpha}$ and $\widehat{\beta}$ we numerically invert the non-linear functions $\nu_{\alpha}$ and $\nu_{\beta}$. This can be done efficiently through a look up table provided for numerous combinations of $\alpha$ and $\beta$ and provided in tabulated form in \citet{McCulloch1986}.}
	\item{\textbf{Obtain estimates of $\gamma$ given estimates of $\alpha$ and $\beta$ using quantile matching:} in \citet{McCulloch1986} a third non-linear function which is explicit in $\gamma$ and implicit in $\alpha$ and $\beta$ is provided in terms of the quantiles as detailed in Equation \eqref{Eqn:McCulloch862}.
\begin{equation}\label{Eqn:McCulloch862}
\nu_{\gamma}(\alpha,\beta)=\frac{x_{(0.75)}-x_{(0.25)}}{\gamma},
\end{equation}
an estimate then follows given $\widehat{\alpha}$, $\widehat{\beta}$ and consistent sample quantiles $\widehat{x}_{(0.75)},\widehat{x}_{(0.25)}$.
}
\end{enumerate}

\subsubsection{Generalized extreme value parameter estimation}\label{subsect:GEV}
A number of methods have been proposed in the literature for estimating parameters in the generalized extreme value family. Here we focus on two such methods, MLE for cases where $\gamma < 0.5$ in \citet{prescott1980maximum} and \citet{Smith1985} and the method of L-moments from \cite{Hosking1990}. We develop a mixed approach combining MLE and L-moments. A detailed discussion on L-moments can be found in \citet{hosking1990moments}, however it is worth noting that in the context of the large data sets utilized for this study, a choice of a mixed method overcomes the instability found in the MLE parameter estimators \citep{Hosking1990}. In addition, the sample L-moments are numerically stable and robust when L-skewness and L-kurtosis estimators are used directly in L-moments estimation. It has also been observed by several authors \citet{royston2006measures} and \citet{vogel1993moment} that the L-moments are less sensitive to outlying data values. The recent developments of mixed methods for inference provide greater computational efficiency, statistical accuracy and robustness in the estimation, which is critical for a high frequency trading strategy that may utilize these methods.

A detailed discussion of the mixed MLE and L-moments based estimation, including asymptotic properties can be found in \citet{Morrison2002}. To present the mixed MLE and L-moments based approach, we first define the L-moments, given by \citet{Hosking1990}, for the real valued random variable $X$ with distribution $F(x)$ and quantile function $Q(p)$ according to Definition \ref{DefnLmoments}.

\begin{definition}[L-Moments]{\label{DefnLmoments} The Population L-moments of a real valued random variable $X \sim F(x)$, for which there is a $K$ sample realization with order statistics given by $X_{(1,K)} \leq X_{(2,K)} \leq \ldots \leq X_{(n,K)} \leq \ldots \leq X_{(K,K)}$, is defined as
\begin{equation}
\lambda_r = r^{-1} \sum_{k=0}^{r-1} (-1)^k \left(\frac{r-1}{k}\right) \mathbb{E}\left[X_{(r-k,r)}\right], \; \forall r = 1,2,\ldots
\end{equation}}
\end{definition}

In practice for such a mixed approach the parameter range of the extreme value index is restricted to $\gamma \in [-0.5,0.5]$ to ensure the moments are finite and appropriate regularity conditions for the MLE are satisfied, see discussion on such items in the generalized Pareto distribution setting and the generalized extreme value setting in \citet[Chapter 5]{Beirlant2004}. We focus on the simplest mixed approach based on the MLE for $\gamma$ and L-moments for $\mu, \sigma$ detailed for a sample size of $K$ according to the following stages.
\begin{enumerate}[Step 1]
\item {\textbf{Re-parametrize the generalized extreme value model likelihood in terms of the extreme value index $\gamma$:} express the parameters $\mu$ and $\sigma$ as functions of $\gamma$ via constraints on the population L-moments given by
\begin{equation}
\begin{split}
\lambda_1 = \mathbb{E}[X_{(1,1)}] &= \mu - \frac{\sigma}{\gamma}\left(1 - \Gamma(1-\gamma) \right),\\
\lambda_2 = \frac{1}{2}\mathbb{E}\left[X_{(2,2)} - X_{(1,2)} \right] &= -2\frac{\sigma}{\gamma}\left(1 - 2^{\gamma}\right)\Gamma(1-\gamma).\\
\end{split}
\end{equation}
}
\item {\textbf{Estimate the population L-moments empirically via the sample L-moments:} which utilizes the ordered data realizations $\left\{x_{n,K}\right\}_{n \in \left\{1,2,\ldots,K\right\}}$ given by
\begin{equation}
\begin{split}
\widehat{\lambda}_1 =\frac{1}{K}\sum_{n=1}^K x_{(n,K)}, \; \mathrm{and} \; \widehat{\lambda}_2 =\frac{1}{K(K-1)}\sum_{n=1}^K \left\{ {{n-1}\choose{1}}  - {{K-n}\choose{1}} \right\} x_{(n,K)}.\\
\end{split}
\end{equation}
Then utilize these estimates to obtain the estimators for $\mu$ and $\sigma$ with respect to $\gamma$ according to the expressions incorporating these L-moment estimates given by
\begin{equation}
\begin{split}
\widehat{\sigma} &= -\frac{1}{2}\gamma \widehat{\lambda}_2\left[\left(1 - 2^{\gamma}\right)\Gamma(1-\gamma)\right]^{-1},\\
\widehat{\mu} &= \widehat{\lambda}_1 + \frac{1}{\gamma}\left(1 - \Gamma(1-\gamma) \right)\left\{-\frac{1}{2}\gamma \widehat{\lambda}_2\left[\left(1 - 2^{\gamma}\right)\Gamma(1-\gamma)\right]^{-1}\right\}.
\end{split}
\end{equation}
}
\item {\textbf{Perform Maximum Likelihood Estimation for the extreme value index parameter $\gamma$ subject to the constraints on L-moments imposed by the estimates in Stage 2:} the maximization of the re-parametrized likelihood given for $\gamma \neq 0$ by
\begin{equation}
\begin{split}
&\ln l(\bm{x}_{(1:K,K)};\widehat{\mu},\widehat{\sigma},\gamma) \approx -K\ln\left(-\frac{1}{2}\gamma \widehat{\lambda}_2\left[\left(1 - 2^{\gamma}\right)\Gamma(1-\gamma)\right]^{-1}\right) \\
& \; - \left(1+1/\gamma\right) \sum_{n=1}^K\ln\left[1+\gamma\mathcal{S}\left(x_{(n,K)},\gamma\right)\right] - \sum_{n=1}^K  \left[1+\gamma\mathcal{S}\left(x_{(n,K)},\gamma\right)\right]^{-1/\gamma},
\end{split}
\end{equation}
where the function $\mathcal{S}\left(x_{(n,K)},\gamma\right)$ is defined according to
\begin{equation}
\mathcal{S}\left(x_{(n,K)},\gamma\right) = \left(\frac{x_{(n,K)}-\left( \widehat{\lambda}_1 + \frac{1}{\gamma}\left(1 - \Gamma(1-\gamma) \right)\left\{-\frac{1}{2}\gamma \widehat{\lambda}_2\left[\left(1 - 2^{\gamma}\right)\Gamma(1-\gamma)\right]^{-1}\right\} \right)}{-\frac{1}{2}\gamma \widehat{\lambda}_2\left[\left(1 - 2^{\gamma}\right)\Gamma(1-\gamma)\right]^{-1}}\right).
\end{equation}
If $\gamma$ is in the neighborhood of the origin $(\gamma \in \mathrm{n.e.}(0))$ the likelihood is given according the Gumbel limit of the generalized extreme value distribution, specified as
\begin{equation} \label{LHGEV_Model_EVI_ZRep}
\begin{split}
&\ln l(\bm{x}_{(1:K,K)};\mu,\sigma,\gamma) \approx -K\ln\left(-\frac{1}{2}\gamma \widehat{\lambda}_2\left[\left(1 - 2^{\gamma}\right)\Gamma(1-\gamma)\right]^{-1}\right) \\ 
& \; - \sum_{n=1}^K\mathcal{S}\left(x_{(n,K)},\gamma\right) - \sum_{n=1}^K  \exp\left[-\mathcal{S}\left(x_{(n,K)},\gamma\right)\right].
\end{split}
\end{equation}
}
\end{enumerate}

\subsubsection{Generalized Pareto distribution parameter estimation}\label{subsect:GPD}
We consider three approaches to estimate the parameters of the generalized Pareto distribution: re-parametrized MLE estimation; analytic Pickands' estimators and robust versions of the empirical percentile method. The re-parametrization of the MLE makes the procedure numerically more robust. 

We consider intra-daily data for each asset during 2010 at time resolutions of 1, 2, 5 and 10 seconds, with trading days split into blocks and utilizing a peaks over threshold approach. Under the peaks over threshold approach the last volume recorded in the specified time increment is retained only if it exceeds the threshold. Comparing this to the preparation of data in Sections \ref{subsect:aSTAB} and \ref{subsect:GEV}, we can see that the $\alpha$-stable estimation considers the full sub-sampled intra-day data sets, generalized extreme value considers the maximum volume per sub-sample time increment and generalized Pareto distribution considers only the largest percentage of volume defined by the specified threshold. 

The section below details the re-parametrized MLE method for estimation of the generalized Pareto distribution model parameters and also the method of moments based estimators for the generalized Pareto distribution parametrization we consider in the paper.

Under the assumption that the volumes collected from the exceedence data are i.i.d. in the peaks over threshold approach, the likelihood for the generalized Pareto distribution as a function of the absolute exceedence data is given for the case in which $1 + \frac{\gamma Y_j}{\sigma} > 0$, by
\begin{equation}\label{EqnLHGPD1}
\begin{split}
&\ln l(\bm{Y};\gamma,\sigma) = -J \log \sigma - \left(\frac{1}{\gamma} + 1\right)\sum_{j=1}^J \log\left(1 + \frac{\gamma Y_j}{\sigma}\right),
\end{split}
\end{equation}
where the condition $1 + \frac{\gamma Y_j}{\sigma} > 0$ ensures the log likelihood is finite. 

If $\gamma = 0$, the likelihood is given according to the exponential based distribution,
\begin{equation} \label{EqnLHGPD2}
\begin{split}
&\ln l(\bm{Y};0,\sigma) = -J \log \sigma - \frac{1}{\sigma}\sum_{j=1}^J Y_j.
\end{split}
\end{equation}

Given the likelihood and the moments of the generalized Pareto distribution distribution or the quantile function, there are numerous statistical approaches we could adopt to perform the parameter estimation. First we discuss how to perform maximum likelihood estimation for such models. Maximization of the generalized Pareto distribution likelihood provided in Equations \eqref{EqnLHGPD1} and \eqref{EqnLHGPD2} with respect to the parameters $\gamma$ and $\sigma$ is subject to the constraints: 
\begin{enumerate}
\item{$\sigma > 0$,}
\item{$1 + \gamma y_{(J)}/\sigma > 0$ where $y_{(J)} = \max\left\{y_1,y_2,\ldots,y_J\right\}$.}
\end{enumerate}
This second constraint is important since one observes that if $\gamma < -1$ as $-\sigma/\gamma \rightarrow y_{(J)}$ then the likelihood approaches infinity. Hence, to obtain maximum likelihood parameter estimates, one should maximize the likelihood subject to these constraints and $\gamma \geq -1$. It is well known that one can in fact re-parametrize the generalized Pareto distribution likelihood to aid in the numerical stability of the parameter estimation via an MLE approach. A re-parametrized MLE version is detailed in \ref{SubSectGPDMLE}, along with the generalized Pareto distribution estimation of parameters via method of moments which is analytic for the parametrization we consider.

\subsubsection{Re-parametrization of the generalized Pareto distribution log-likelihood and maximization} \label{SubSectGPDMLE}

In practice it is beneficial to consider a re-parametrization of the generalized Pareto distribution log-likelihood function according to 
\begin{equation}
\left(\gamma,\sigma\right) \rightarrow \left(\gamma,\tau\right) = \left(\gamma, \frac{\gamma}{\sigma} \right),
\end{equation}
producing a re-parametrized log-likelihood model given by
\begin{equation}
\ln l(\bm{Y};\gamma,\tau) = -J \ln\gamma + J \ln \tau - \left(\frac{1}{\gamma}+1\right)\sum_{i=1}^J \ln\left(1 + \tau Y_i \right).
\end{equation}
This log-likelihood is then maximized subject to $\tau < 1/y_{(J)}$ and $\gamma \geq -1$. Under the first partial derivative this produces, 
\begin{equation}
\begin{split}
&\frac{\partial \ln l(\bm{Y};\gamma,\tau)}{\partial \gamma} = 0;\\
&\Rightarrow \; \gamma(\tau) = \frac{1}{J}\sum_{j=1}^J \ln(1 - \tau y_j).
\end{split}
\end{equation}
Hence, the estimation is performed in two steps:
\begin{enumerate}
\item{Estimate $\widehat{\tau}^{MLE} = \arg \max \ln l(\gamma(\tau),\tau)$ subject to $\tau < 1/y_{(J)}$.}
\item{Estimate $\widehat{\gamma}^{MLE} = \frac{1}{J}\sum_{j=1}^J \ln(1 - \widehat{\tau}^{MLE} y_j)$. Then solve for the original parameterization via inversion $\widehat{\sigma}^{MLE} = \frac{-\widehat{\gamma}^{MLE}}{\widehat{\tau}^{MLE}}$.}
\end{enumerate}
Note that the log likelihood $\ln l(\gamma(\tau),\tau)$ is continuous at $\tau=0$, hence if the estimator $\widehat{\tau}^{MLE}=0$ then one should consider $\widehat{\gamma}^{MLE}=0$ and 
\begin{equation}
\widehat{\sigma}^{MLE} = \frac{1}{J}\sum_{j=1}^J y_j.
\end{equation}
In addition, in practice to ensure that $\gamma \geq 1$, the condition that $\tau < 1/y_{(J)}$ should be modified to $\tau < (1-\epsilon)/y_{(J)}$, where $\epsilon$ is found from the condition that $\gamma(\tau) \geq -1$.

\begin{remark}{It has been shown in \citet{Smith1985} and \citet{Embrechts1997} [Section 6.5.1] that in the case in which $\gamma > -1/2$ the MLE vector $\left(\widehat{\gamma}^{MLE},\widehat{\sigma}^{MLE}\right)$ is asymptotically consistent and distributed according to a bivariate Gaussian distribution with asymptotic covariance, obtained using the MLE parameter estimates given by the usual Fisher information matrix. When the data is not independent the results will continue to be consistent asymptotically normal, but the asymptotic covariance matrix is substantially more complicated than the independent case.} 
\end{remark}

\subsubsection{Empirical percentile method} \label{SubSectGPDepm}

The Empirical percentile method approach to parameter estimation is based on the percentile based matching approach proposed in \citet{Castillo1997}. We equate the model CDF evaluated at the observed order statistics to their corresponding percentile values. This system of equations can then be solved for the models distributional parameters. In the case of the generalized Pareto distribution model for the volume profiles on the bid and ask at level 1, there are two model parameters, so we require two distinct order statistics as a minimum to perform the estimation. 

Consider a set of realized data obtained under a peaks over threshold approach, where the $J$ volumes that have exceeded a pre-specified threshold level $u$ are denoted by the data $\left\{x_i\right\}_{i=1:J}$ with order statistics denoted by $\left\{x_{(i,J)}\right\}_{i=1:J}$. 

Given the CDF of the generalized Pareto distribution model in Equation \eqref{EqnGPDCDF}
\begin{equation} \label{EqnGPDCDF}
F(x;\gamma,\sigma) = \begin{cases}
1 - \left(1 - \frac{\gamma x}{\sigma}\right)^{\frac{1}{\gamma}}, \; \gamma \neq 0, \sigma > 0,\\
1 - \exp\left(-\frac{x}{\sigma}\right), \; \gamma = 0, \sigma > 0,
\end{cases}
\end{equation}
we match the CDF at two of the selected order statistics $i \neq j \in \left\{1,2\ldots,J\right\}$ to the corresponding percentile values,
\begin{equation}
F\left(x_{(i,J)};\gamma,\sigma\right) = p_{(i,J)} \; \; \mathrm{and} \; \; F\left(x_{(j,J)};\gamma,\sigma\right) = p_{(j,J)},
\end{equation}
where the percentile is given for the generalized Pareto distribution model with $J$ observations by 
\begin{equation}
p_{(i,J)} = \frac{i - \eta}{J+\zeta}.
\end{equation}
It is recommended in \citet{Castillo1997} that choices of $\eta = 0$ and $\zeta = 1$ provide reasonable results, so these settings were utilized in the studies performed. The solution to this system of equations in terms of the parameters is obtained by solving the equations for $\gamma$ and $\sigma$ given by
\begin{equation}
1 - \left(1 - \frac{\widehat{\gamma} x_{(i,J)}}{\widehat{\sigma}}\right)^{\frac{1}{\widehat{\gamma}}} = \frac{i}{J+1} \;\; \mathrm{and} \;\; 1 - \left(1 - \frac{\widehat{\gamma} x_{(j,J)}}{\widehat{\sigma}}\right)^{\frac{1}{\widehat{\gamma}}} = \frac{j}{J+1}.
\end{equation}
Hence for any two pairs of order statistics $i,j$ the solutions to these system of equations is 
\begin{equation}
\widehat{\gamma}(i,j) = \frac{\ln\left(1 - \frac{x_{(i,J)}}{\widehat{\delta}(i,j)}\right)}{C_i}\; \; \; \mathrm{and} \;\; \widehat{\sigma}(i,j) = \widehat{\gamma}(i,j)\widehat{\delta}(i,j), 
\end{equation}
in terms of $C_i = \ln\left(1-p_{(i)}(J)\right) < 0$ and $\widehat{\delta}(i,j)$. Here $\widehat{\delta}(i,j)$ is the solution to the equation,
\begin{equation}
C_i \ln\left(1 - \frac{x_{(j,J)}}{\delta}\right) = C_j \ln\left(1 - \frac{x_{(i,J)}}{\delta}\right),
\end{equation}
which is obtained using a univariate root finding algorithm such as bisection. Note, that $\delta$ corresponds to a re-parametrization of the generalized Pareto distribution distribution when $\delta = \frac{\sigma}{\gamma}$.

\begin{remark}[Empirical percentile method and Pickands' analytic solution]{A special case of the empirical percentile method estimators is widely used in estimation of the generalized Pareto distribution model parameters and known as the Pickands' estimator. These correspond to the empirical percentile method setting in which $i = \frac{J}{2}$ and $j = \frac{3J}{4}$. In these special cases, the bisection method is not required as the system of equations can be solved analytically according to
\begin{equation} 
\widehat{\gamma} = \frac{1}{\ln 2}\ln\left(\frac{x_{(J/2,J)}}{x_{(3J/4,J)}-x_{(J/2,J)}}\right) \;\; \mathrm{and} \;\; \widehat{\sigma} = \widehat{\gamma}\left(\frac{x^2_{(J/2,J)}}{2x_{(J/2,J)}-x_{(3J/4,J)}}\right).
\end{equation}
}
\end{remark}

In general we would not just pick two indexes $i,j$ and instead we would combine the Algorithm's one and two discussed in \citet{Castillo1997} to produce an estimate of the generalized Pareto distribution parameters. Combining Algorithm 1 and Algorithm 2 of \citet{Castillo1997} we follow the stages outlined below to estimate the generalized Pareto distribution parameters via the empirical percentile method.
\begin{enumerate}[Step 1]
\item{\textbf{Repeat the following steps for all order indexes $\left\{i,j: i<j, \text{ for } i,j \in \left\{1,2,\ldots,J\right\} \right\}$, such that $x_{(i,J)} < x_{(j,J)}$:}}
\begin{enumerate}
\item{Compute for $s \in \left\{i,j\right\}$ the values $C_s = \ln\left(1-\frac{s - \eta}{J+\zeta}\right)$.}
\item{Set $d = C_j x_{(i,J)} - C_i x_{(j,J)}$, if $d = 0$ let $\widehat{\delta}(i,j) = \pm \infty$ and set the EVI estimate $\widehat{\gamma}(i,j) = 0$, otherwise compute $\delta_0 = x_{(i,J)}x_{(j,J)}(C_j - C_i)/d$.
}
\item{If $\delta_0 > 0$, then $\delta_0 > x_{(j,J)}$ and the bisection method can be used for the interval $[x_{(j,J)}, \delta_0]$ to obtain a solution $\widehat{\delta}(i,j)$. Otherwise the bisection method is applied to the interval $[\delta_0,0]$.}
\item{Use $\widehat{\delta}(i,j)$ to compute $\widehat{\gamma}(i,j)$ and $\widehat{\sigma}(i,j)$ using
\begin{equation}
\begin{split}
\widehat{\gamma}(i,j) &= \frac{\ln\left(1 - \frac{x_{(i,J)}}{\widehat{\delta}(i,j)}\right)}{C_i} \; \mathrm{and} \; \widehat{\sigma}(i,j) = \widehat{\gamma}(i,j)\widehat{\delta}(i,j).
\end{split}
\end{equation}
}
\end{enumerate}
\item{\textbf{Take the median of each of the sets of estimated parameters for the overall estimator to obtain:}
\begin{equation}
\begin{split}
\widehat{\gamma}^{EPM} &= \text{median}\left\{\widehat{\gamma}(1,2),\widehat{\gamma}(1,3),\ldots,\widehat{\gamma}(J-1,J)\right\},\\
\widehat{\sigma}^{EPM} &= \text{median}\left\{\widehat{\sigma}(1,2),\widehat{\sigma}(1,3),\ldots,\widehat{\sigma}(J-1,J)\right\}.
\end{split}
\end{equation}
}
\end{enumerate}

\subsection{Goodness-of-fit testing for heavy tailed models of LOB volume profiles}\label{sect:param est}
In this section we present a class of omnibus compound goodness of fit hypothesis testing procedures specifically designed for heavy tailed models.

To assess the quality of the statistical model estimations, we considered a number of approaches. The first involves an exploration of the goodness of fit utilizing the Kolmogorov-Smirnov statistical test (classical Kolmogorov-Smirnov test) which aims to test the compatibility of the theoretical probability distribution with the empirical probability distribution. The classical versions of the omnibus goodness of fit test based on the Kolmogorov-Smirnov supremum statistic proved inappropriate for two reasons: the first is it inadequately assesses the key feature we consider namely the heavy tails; secondly the massive data sets tend to result in each hypothesis being tested under thousands of samples resulting in criteria so strict that large rejections will arise unless the test is directly targeting the appropriate null. Since the test is not correctly attributing appropriate weights to the sub-exponential tails of the model, the test will incorrectly reject the null as the sample increase at a disproportionate rate to reality \citep{Chickeportiche2012}, which is what we observed in the testing of all distributions. This is due primarily to the test expecting an exponential rather than a power law decay in the tail probabilities, see detailed discussion in \citet{koning2008goodness}.

Due to the inappropriateness of the classical Kolmogorov-Smirnov test for heavy tailed, we implement the variance weighted modified version of the Kolmogorov-Smirnov test discussed in \citet{Chickeportiche2012}. To implement the modified Kolmogorov-Smirnov test, consider i.i.d. random variables $\left\{X_i\right\}_{i=1}^N$ with distribution $F$. We let $Y_n(x)=\mathbb{I}_{\{X_n\leq x\}}$, of which the components are Bernoulli variables. The centered sample mean measures the difference between the empirical CDF and the theoretical CDF at point $x$. We define the centered sample mean, $Y(x)$ as:
\begin{equation}
	\overline{Y}(x)=1/N\sum_{n=1}^{N} Y_n(x) - F(x),
\end{equation}
where $N$ is the sample size. Let $u=F(x)$ and thus,
\begin{equation}
	\overline{Y}(u)=1/N\sum_{n=1}^{N} \sqrt{N} Y_n(F^{-1}(u)) - u.
\end{equation}
The variance weighted Kolmogorov-Smirnov test then has equi-weighted quantiles, which is equally sensitive to all regions of the distribution, including importantly the tails. The resulting weighting is defined as,
\begin{equation}
	\tilde{y}=y(u)\sqrt{\phi(u;a,b)},
\end{equation}
where,
\[ \phi(u;a,b) = \left\{\begin{array}{cc} \frac{1}{u(1-u)}, & a\leq u \leq b,\\
																				0, & \text{otherwise.} \end{array} \right. \]
The choices for $a$ and $b$ are $a=1/N$ and $b=1-a$, corresponding to the minimum and maximum of the sample (as suggested by \citet{Chickeportiche2012}). It follows that we evaluate the variance weighted supremum test statistic according to
\begin{equation}
	K(a,b) = \sup_{u\in[a,b]}|\tilde{y}(u)|=\left\{\begin{array}{cc} \frac{y(u)}{\sqrt{u(1-u)}}, & a\leq u \leq b,\\
																				0, & \text{otherwise.} \end{array} \right.
\end{equation}
Then, under the null the critical threshold value $k^*$ corresponding to a 95\% confidence level is determined from the following equation. 
\begin{equation}
\frac{-\ln 0.95}{\ln N} \approx \sqrt{\frac{2}{\pi}}k^{*}\exp{-\frac{k*^2}{2}}.
\end{equation}
We implement a root finding method to determine $k^{*}$, which is dependent on the sample size N. 

\section{Results and discussions on model estimation and goodness of fit for LOB volume profiles}\label{Results2}
Previous studies have failed to capturing the features presented in Section~\ref{sec:Results1} when building statistical models for LOB volumes. This has direct ramifications on both high frequency trading strategies and achieving best execution practices. 

The purpose of this section is to formally study the features of the volume profiles of level 1 bid and level 1 ask in the LOB for futures markets using more flexible families of parametric models. We firstly, assess the appropriateness of statistical models and fit at different sampling frequencies. We then consider the appropriateness of a heavy tailed assumption for the volume profiles each day as captured by the sub-exponential family models for volume profile tails given by: \textsl{$\alpha$-stable, generalized Pareto distribution and generalized extreme value distributional families}. In the process of estimating these models for the LOB volume profile data, we also assess, study and recommend sophisticated statistical estimation procedures and their performance for each model for practitioners when estimating such models for the large volume of data in a LOB model analysis which included: \textsl{MLE; Generalized method of moments and L-moment estimation; Mixed methods of MLE and generalized moment matching; Empirical percentile estimation and Quantile based estimators}. 

\subsection{$\alpha$-stable model estimation results for LOB volume profiles}\label{sec:aStabModRes}
The $\alpha$-stable family of distributions were fitted to volume data on level one of the bid and ask, scaled by the inter quartile range. For each day of data we utilized McCullochs method to estimate the parameters. We begin with analysis of the dynamics of the parameters that define the $\alpha$-stable distribution across the 6 assets analyzed, 5YTN, BOBL, SP500, NIKKEI, GOLD and SILVER. 

We first recall that for the $\alpha$-stable distribution, the tail index $\alpha \in (0, 2]$ determines the rate of tail decay and is the key parameter of interest. The parameter $\alpha$ is typically termed the characteristic exponent, with small and large $\alpha$ implying heavy and light tails respectively. Gaussian ($\alpha = 2, \beta = 0$) and Cauchy ($\alpha = 1, \beta = 0$) distributions provide the only simple analytically tractable sub members of this family. 

For the majority of 2010, the 5YTN had tail index parameter estimates indicative of light tails for volumes on the bid and ask, with a mean of 1.9895. However, a particularly interesting feature involved a few pronounced periods in which a heavy tail model is clearly appropriate for the volume profile. The days on which these heavy tailed volume profiles occurred did not correspond to the same days for the bid and ask volume profiles, indicating an asymmetry in the volume profile on the bid and ask over time, with respect to extreme volumes. Additionally, volumes for the 5YTN also appear to be heavily right skewed with a $\beta \approx 1$ a large portion of the year. 

\begin{figure}[h]
\begin{center}
\begin{minipage}{150mm}
{\includegraphics[height = 7.5cm, width = \textwidth]{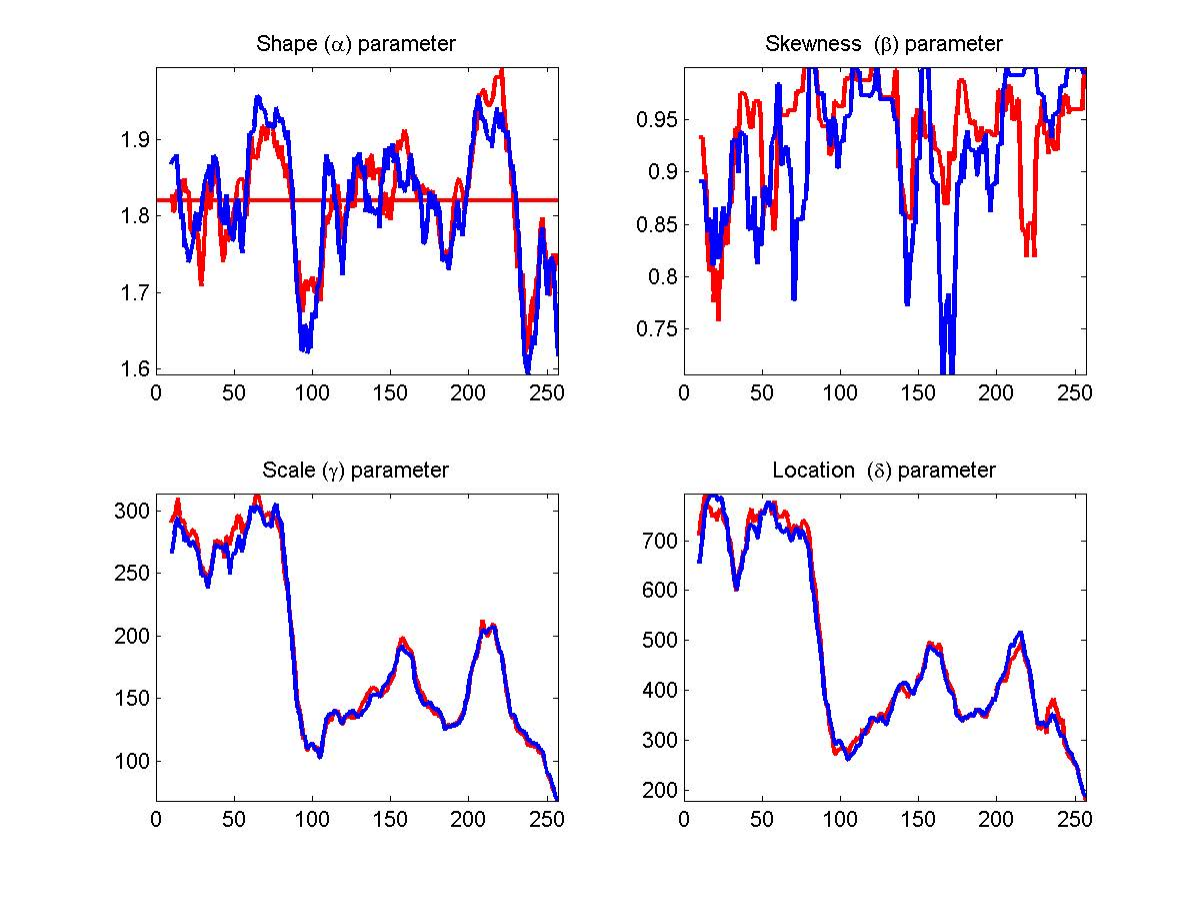}
  \caption{$\alpha$-stable 10-day moving average of the daily parameter estimation for the year 2010 using McCullochs method for \textbf{BOBL} at a time resolution of 10 seconds. The blue line is the bid L1 and red line ask L1. \textbf{Top Left Plot:} Tail index parameter $\alpha$ daily estimates. \textbf{Top Right Plot:} Asymmetry parameter $\beta$ daily estimates. \textbf{Bottom Left Plot:} Scale parameter $\gamma$ daily estimates. \textbf{Bottom Right Plot:} Location parameter $\delta$ daily estimates.}
\label{fig:AStabBOBLPar}}
\end{minipage}
\end{center}
\end{figure}

In Figure \ref{fig:AStabBOBLPar} we contrast the daily estimated findings for the 5YTN to BOBL. We can see a significantly greater variation in the tail index $\alpha$ and skewness parameter $\beta$, with the mean of the tail index parameter being 1.8195. This indicates that the daily volume profile on the bid and ask is consistently more heavy tailed than the daily behavior of the volume profile for the 5YTN. In addition, the extreme volume profile events observed occasionally in the 5YTN are not present in the BOBL volume profile, until the end of 2010 where an event caused the volume profile to demonstrate an infinite mean tail behavior for a few trading days in late November to early December. This is consistent with the observed extreme events estimated from the $\alpha$-stable daily model fits during this period.

\FloatBarrier

The daily volume profiles for the LOB for the NIKKEI were similar in nature to the BOBL's consistent heavy right tail attributes with strong skew and a mean tail index of 1.8022. In addition, we observed for the NIKKEI that the daily volume profile on the bid and ask is not only consistently heavy tailed but asymptotically dominates the behavior in the volume profile of BOBL and 5YTN. Whilst there is symmetry between the bid and ask sides, there are some exceptions and in these exceptions there is a marked relationship between asymmetry in the bid and ask volume profiles with the bid tending to produce a symmetric distributional fit when the ask is asymmetric and vice versa.
 
Comparing the SP500 with the other assets considered, we see an estimated tail index parameter $\alpha$ that is more pronounced, with a mean of 1.7474. In addition, we see a consistent daily tail profile which had a tail index away from $\alpha < 2$. What is also of interest here is that when considering the SP500, which was globally the 2nd most actively traded equity future in 2010, there was actually a total volume decrease between 2009 and 2010 of $-3.0\%$. However, this total change in volume did not affected the general attributes observed for the model estimation with regards to the heavy tailed behavior and the manner in which these contracts are traded on an intra-daily basis throughout the year.
 
For the precious metals explored, the most prominent of the heavy-tailed volume profiles is the GOLD futures which had a mean tail index parameter $\alpha$ of 1.5076 and consistently heavy tailed behavior intra-daily throughout all trading days in 2010. GOLD also demonstrated a strongly right skewed distribution for the volume profile at level 1 of the bid and ask. The results for SILVER demonstrated a few marked periods in which the intra-daily volume profile on both the bid and ask became exceptionally heavy tailed in nature, most noticeably in the mid-year in which the ask side had tail index values around $\alpha \approx 1$. 

\subsection{Generalized extreme value model estimation results for LOB volume profiles}
Consistent with the findings from the $\alpha$-stable model, the estimation results for the 5YTN show heavy-tailed behavior is present in the volume profiles between the 50th and 60th trading day and the 140th and 150th trading days. Interestingly, the prominence of the heavy tailed features for BOBL is more pronounced under the generalized extreme value model fits compared to the $\alpha$-stable model. Additionally, the structural changes in the behavior of the intra-daily volume profile on the bid and ask are observed in the location and scale parameters for the BOBL around the 100th trading day, where there is a marked regime shift in the estimated model parameters. This is just as prominent in the generalized extreme value fit as it was in the $\alpha$-stable model, indicating that it is entirely plausible that there was a dynamic change in the intra-day activity in this market mid trading year. A similar change is visible in the SP500 shape and scale parameters, but in this case the regime reverts gradually back to the behavior present intra-day at the start of the year. This structural change is not as prominent in the NIKKEI.

\begin{figure}[h]
\begin{center}
\begin{minipage}{150mm}
{\subfigure[MLE approach.]{
\resizebox*{7.5cm}{!}{\includegraphics{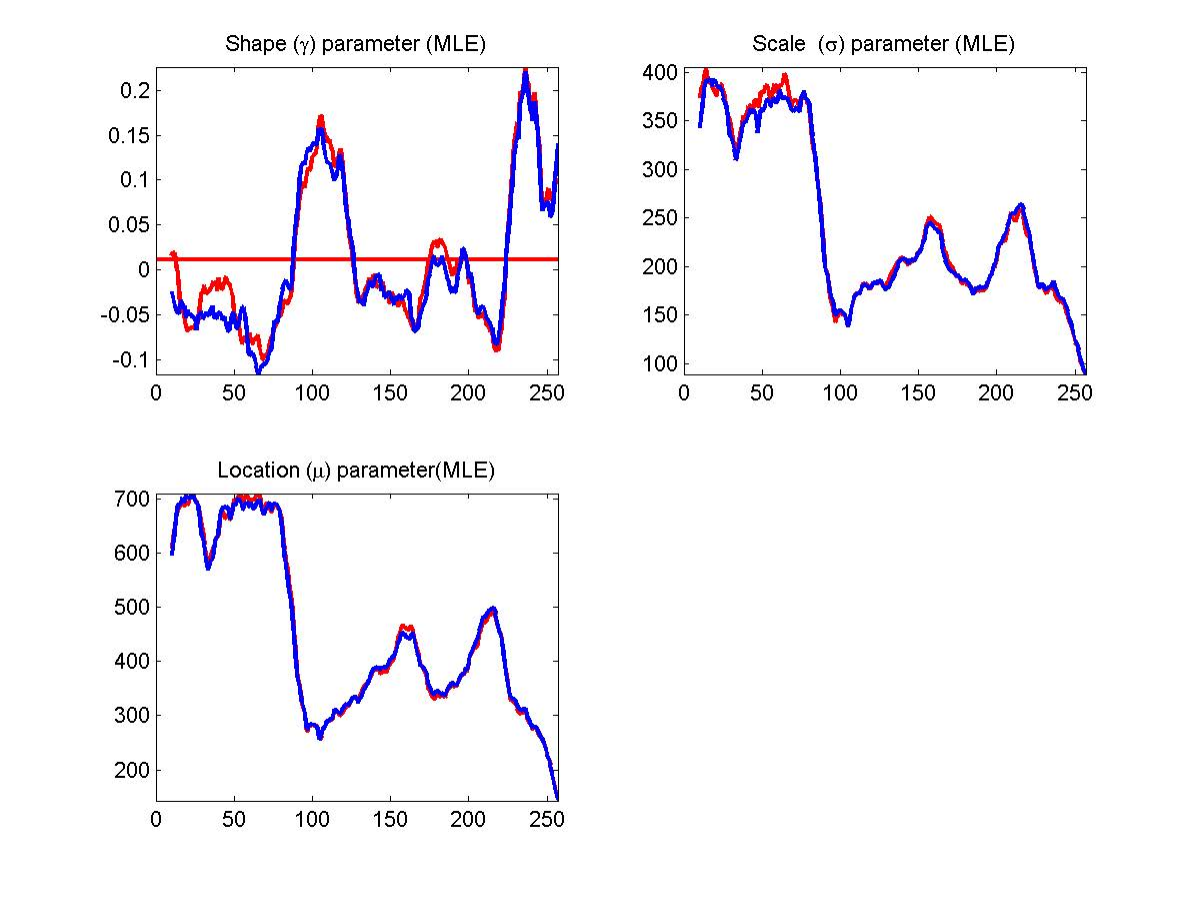}}\label{sample-figure_part_a}}
\subfigure[Mixed L-moments approach.]{
\resizebox*{7.5cm}{!}{\includegraphics{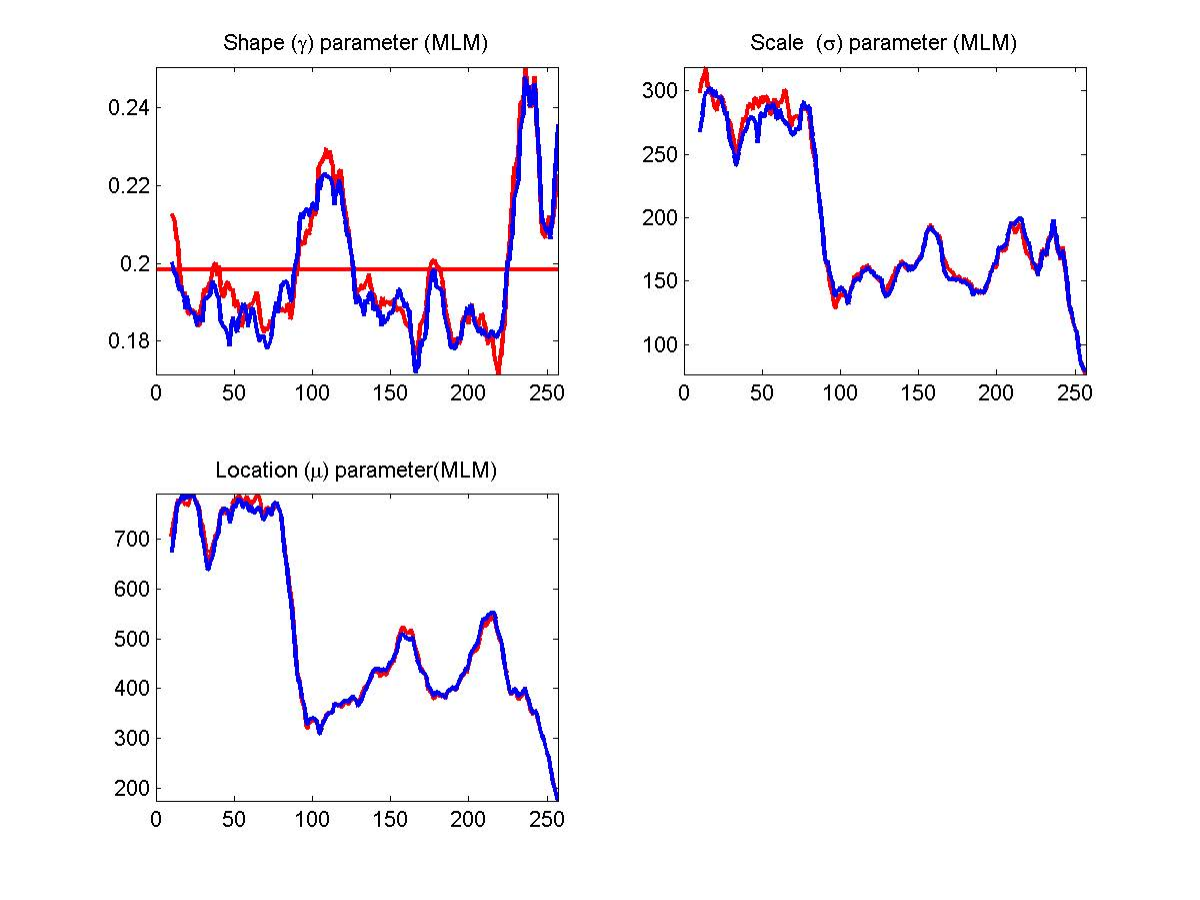}}\label{sample-figure_part_b}}
\caption{Generalized extreme value intra-day 10-day moving average of the parameter estimation on each trading day of the year for BOBL bid and ask side at time resolution of 10 seconds.}
\label{fig:GEVBOBLParMLE}}
\end{minipage}
\end{center}
\end{figure}

When considering the tail index parameter, ($\gamma$) for the MLE method, we can see that the mean intra-daily estimated value averaged over all trading days in 2010 for the 5YTN, BOBL (Figure ~\ref{fig:GEVBOBLParMLE}) and NIKKEI is close to zero, respectively -0.0493, 0.0114, 0.0725. SP500 has a higher mean level for the shape parameter, being 0.1495. Again, as in the $\alpha$-stable case, the generalized extreme value model estimations do indicate a reasonable variation in the tail index throughout 2010, indicating a number of days in which heavy tailed volume profile attributes are appropriate. On a few days analyzed and for assets 5YTN, SP500 and NIKKEI we see instances where the shape parameter spikes ($\gamma>1$) indicating infinite mean-variance models are suitable. The occurrence of such events coincides with the trading days in which the $\alpha$-stable model also indicated infinite mean heavy tailed behavior as suitable.  
For all assets we see a correlation in the parameter estimations for the bid and ask side and all assets show time variation across the year for scale and location.  

The mixed L-moments approach is used to confirm our findings of the estimation of the tail index parameter, which is notoriously difficult to estimate. Comparing the subplots in Figure \ref{fig:GEVBOBLParMLE} for BOBL we have a good representation of the features we found to be consistent across the assets 5YTN, SP500 and NIKKEI. The scale and location parameters are very similar for each method implemented. However we see that the shape parameter is systematically different when comparing the MLE and mixed L-moments methods, with the mean level for the shape parameter using the MLE method being between $(-0.0495, 0.1495)$ for all assets, whereas the mixed L-moments method produces a mean level of $(0.1951, 0.2321)$ for the shape parameter for all assets. 

\FloatBarrier

To further explore the discrepancies between the MLE and mixed L-moments, specifically the upward translation of the shape parameter by approximately $2$, we performed a simulation study which considered sample sizes ranging from 50 to 10,000. We randomly generated generalized extreme value distributed data series and replicate each sample 20 times, estimating the parameters of the generalized extreme value distribution for each replication and each sample size using MLE and mixed L-moments methods. Results show the same discrepancies between the different estimation methods for the shape parameter. As $\gamma \rightarrow 0$, we see an increased bias of $\gamma$ under the mixed L-moments method, but the tradeoff is that the variance is reduced in the mixed L-moments method. Mixed L-moments becomes more reliable as the sample size decreases, thus prompting us to recommend the use of MLE for higher frequency LOB volume data. 
 
In summary, we observed that the MLE method provided a more stable fit compared with the mixed L-moments method for these applications. Interestingly, we found from the analysis that as the data becomes significantly heavier tailed, as was the case for GOLD with a mean intra-day tail index parameter of 0.3564 for 2010, the results for the MLE estimation and the L-moments based solutions were in much closer in alignment. The observed bias present in the cases of light tailed volume profiles on certain trading days in the BOBL was not present in the consistently heavy tailed GOLD.

\subsection{Generalized Pareto distribution model estimation results for LOB volume profiles}\label{sec:GPD}

The generalized Pareto distribution family utilize a peaks over threshold preparation of the data, with a translation by a threshold corresponding to the 80th percentile of the data (the location ($\mu$) parameter), see discussion in \citet{embrechts1999extreme}. The features observed under the generalized pareto distribution model are consistent with the findings discussed for the $\alpha$-stable and generalized extreme value models. In particular the prevalence for the heavy tailed attributes remain, as does the interesting features of the increased extreme intra-day volume activities resulting in heavy tailed attributes appearing for the BOBL, as observed in the generalized extreme value and $\alpha$-stable fits.

The scale parameter, $\sigma$ for the generalized Pareto distribution distribution, using the MLE method for estimation, shows some structural shifts for 5YTN and SP500 around the same time period as the structural shifts observed in the $\alpha$-stable distribution.

\begin{figure}[h]
\begin{center}
\begin{minipage}{145mm}
{\subfigure[MLE approach.]{
\resizebox*{7.5cm}{!}{\includegraphics{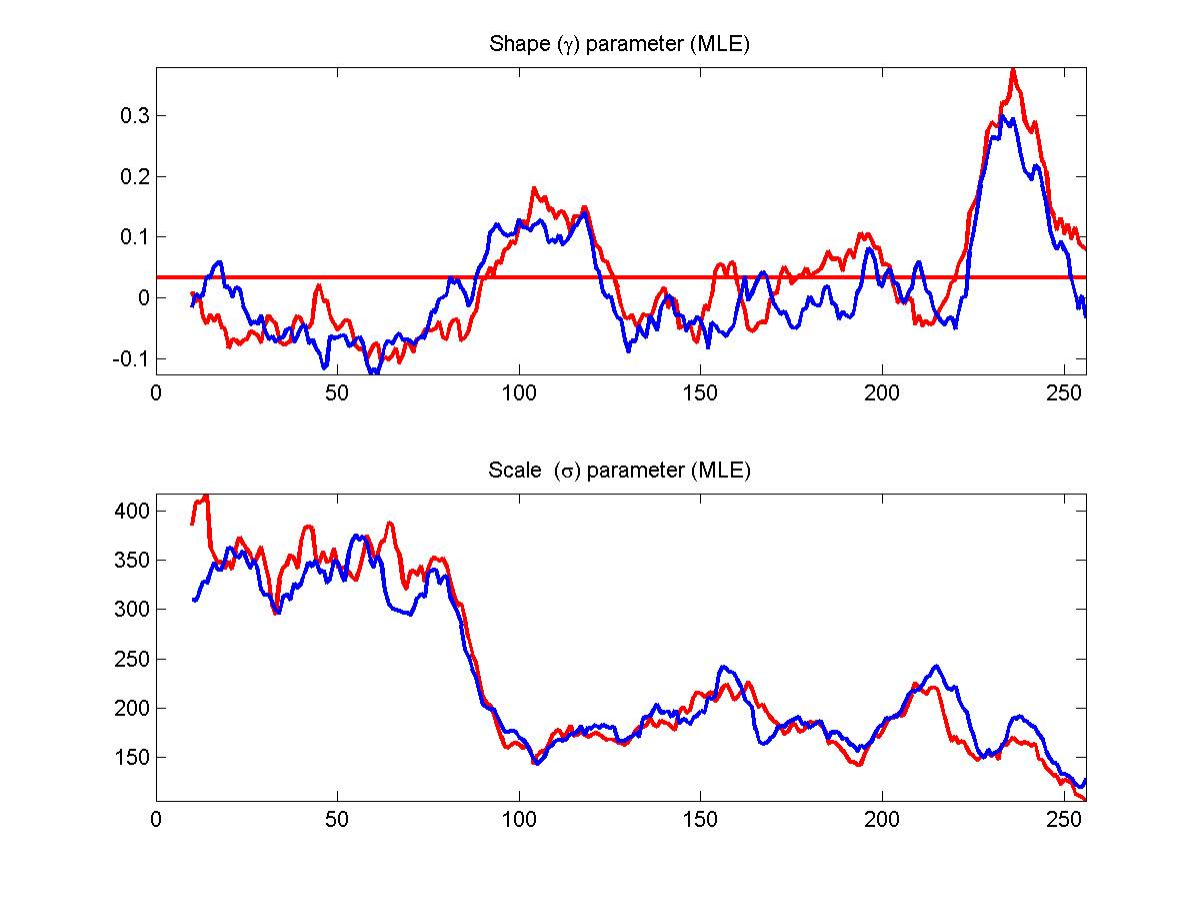}}\label{sample-figure_part_a}}
\subfigure[Pickand method.]{
\resizebox*{7.5cm}{!}{\includegraphics{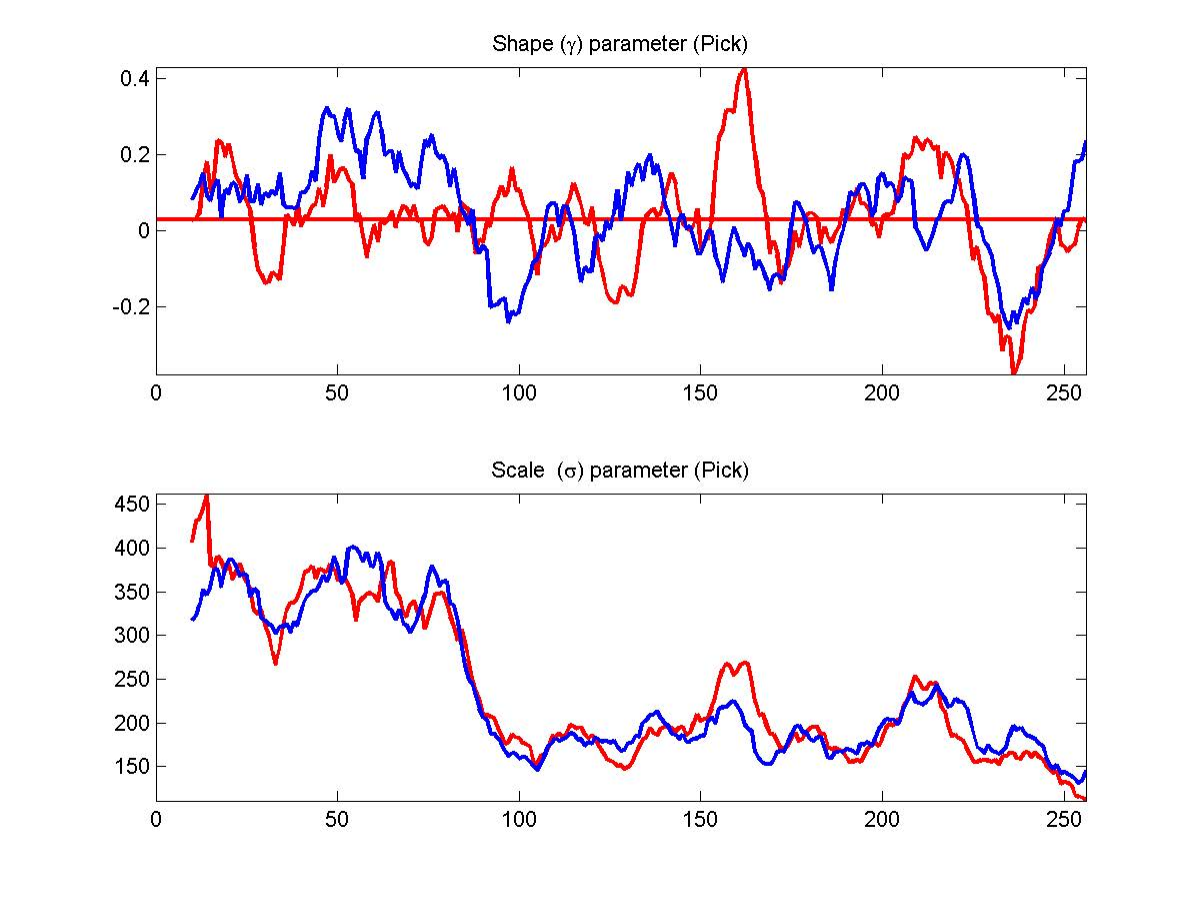}}\label{sample-figure_part_b}}
\caption{Generalized Pareto distribution intra-day 10-day moving average of the parameter estimation for each trading day of 2010 for the BOBL bid and ask side at time resolution of 10 seconds.}
\label{fig:GPDBOBLMLEPick}}
\end{minipage}
\end{center}
\end{figure}
\FloatBarrier

In terms of the different estimation approaches, the scale parameter using the Pickands' estimator shows similar trending features as the MLE method. However, it also exhibits much higher variability in parameter estimation compared with the MLE method. The Pickands' estimator fails to capture the days where we see a significant spikes in the shape and scale parameters. 

The results observed for the empirical percentile method demonstrated a substantial deviation from the MLE and the Pickand's estimator. The implementation of the empirical percentile method that we adopted involved matching each pair of percentiles above a threshold of the median to obtain a solution for the model parameters, producing a very large number of solutions. Next we took the median of these solutions as a robust estimation of the model parameters under the empirical percentile method, as discussed in \cite{Castillo1997}. Our findings indicated the sensitivity of this estimation approach to the inclusion of low percentile solutions into the estimation (median calculation) for the shape (tail index) parameter. 

To further explore why the empirical percentile method gives substantially different results to the MLE and Pickands' methods, we considered a synthetic simulation study with 20 data sets, each with 500 randomly generated GP data points used to estimate the MLE, Pickands' and empirical percentile method. We considered the impact of setting the shape parameter positive and negative. The results show that for a positive shape parameter, the three methods appear to be consistent. However, when the actual shape parameter is negative we see a significant translation upwards, with increased variability in the estimator, consistent with results for the observed three methods when analyzing the real LOB volume data. 

We note the empirical percentile method of estimation has significant issues when attempting to apply this method to real data. We found that the method produces a much more stable result when using higher starting percentiles for the grid search method for maximizing the log-likelihood function. However, for the simulated and real case there seems to be a systematic bias in the estimation under the empirical percentile method but not the MLE when the data appears to have an \textit{actual} negative shape parameter. It should also be noted that the starting percentile used for this method was the 50th percentile.

In general we found little difference in the estimated model parameters at each of the sampling frequencies and on these grounds we proceed with discussion of the 10 seconds sampling rate. The consistent presence of these features at all these sampling rates allows us to conclude that irrespective of whether a high frequency trading strategy targets the ultra-high frequency range or is generally volume sensitive, these features need to be accounted for in LOB liquidity studies, LOB resilience studies and execution studies.The premise of this conclusion is that such activities tend to take place at the high frequencies ($<1$ second).

\subsection{Goodness-of-Fit via variance weighted modified Kolmogorov-Smirnov test}\label{sec:casestudy}
In this section we consider each of the three fitted model estimations daily for 2010. On each day of estimation we perform a goodness of fit analysis using the specifically modified variance weighted Kolmogorov-Smirnov test for each model. Interestingly we observed that when using the modified variance weighted Kolmogorov-Smirnov test we found that the resulting test rejects the null hypothesis of the $\alpha$-stable distribution providing a good fit to the data for all assets. Hence, we conclude that whilst the model was able to be estimated efficiently, the fit is not sufficiently reflective of the data, with rejection of the null hypothesis. The modified Kolmogorov-Smirnov test rejects the null hypothesis of the generalized extreme value distribution providing a good fit to the data for all assets using the mixed L-moments method. However, the MLE method provides a good fit to the data for SP500 22\% of the trading days for the bid side and 17\% of the trading days for the ask side. For BOBL and NIKKEI, we see that it provides a good fit less than 10\% of the trading days on both the bid and ask side. 

\begin{figure}[h]
\begin{center}
\begin{minipage}{150mm}
{\includegraphics[height = 7.5cm, width = \textwidth]{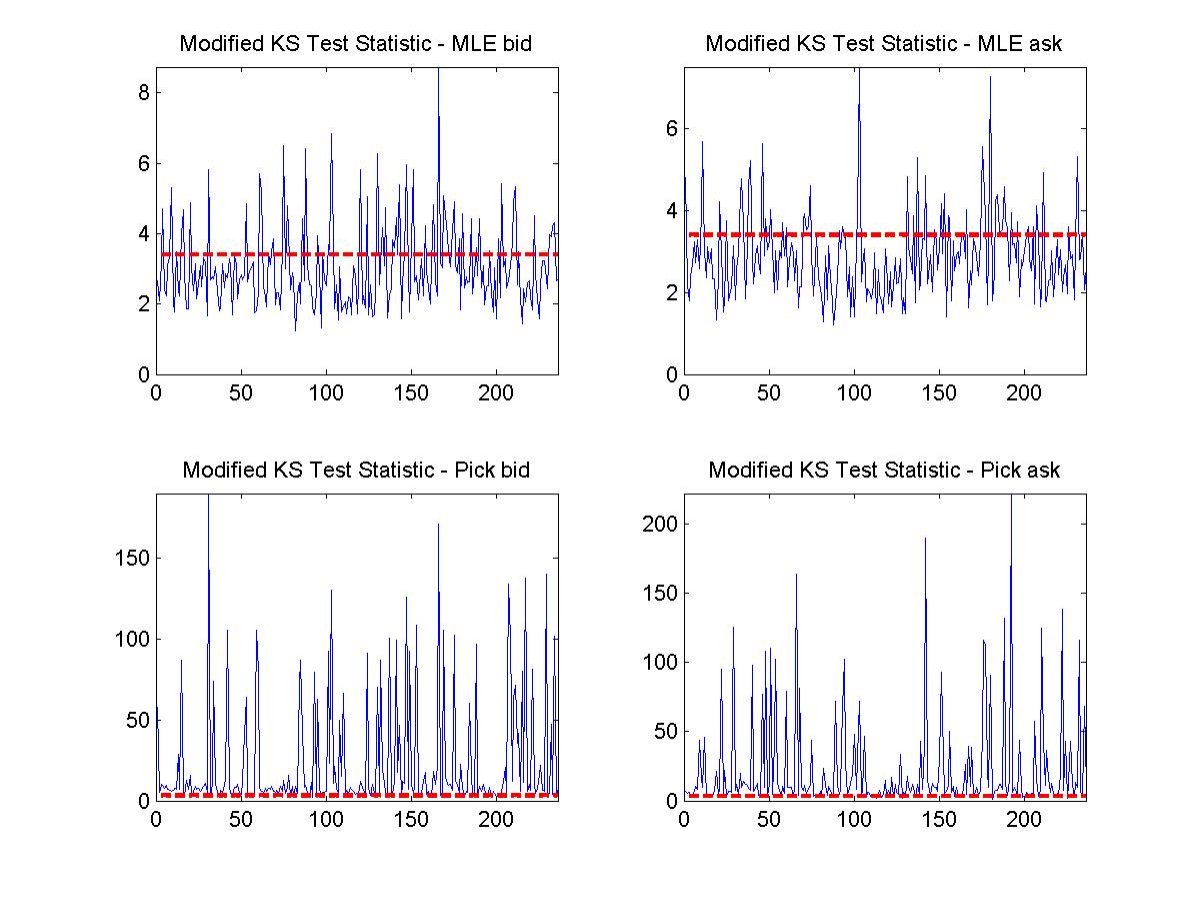}
  \caption{Modified KS-Test statistics for 5YTN for every trading day bid and ask side using generalized pareto distribution MLE and Pickands' method.}
\label{fig:modKSGPD5YTNGoF}}
\end{minipage}
\end{center}
\end{figure}

From figure ~\ref{fig:modKSGPD5YTNGoF} we can see that the modified KS test does not reject the null hypothesis at a 5\% level of significance of the generalized Pareto distribution using the MLE estimation method. For the 5YTN we see that the MLE estimation method provides a good fit for 72\% of trading days for the bid and 76\% of trading days for the ask. Whereas the Pickands' method only provides a good fit for 8\% of trading days on the bid and 6\% of trading days on the ask for the 5YTN. These results are largely consistent with all other assets, however we see that generalized Pareto distribution MLE method provides a good fit for 95\% of days for the bid side and 88\% of days for the ask side for the SP500. Likewise the Pickands' results were also better for the SP500 with a good fit being observed 17\% of trading days on the bid and 10\% on the ask side. We do not present the results for the empirical percentile method for the modified KS test due to the estimation issues outlined above.

\FloatBarrier

We considered 6 different estimation methods across three different distributions, $\alpha$-stable, generalized extreme value and generalized Pareto distribution. Table ~\ref{tab:concdist} shows the superior method for each of the distributions. Across all methods, the generalized Pareto distribution MLE method provided the best fit to the data across all assets. The parameter estimates did not vary significantly with a change in sampling rate and the weighted modified Kolmogorov-Smirnov test concluded the generalized Pareto distribution MLE method provided a good fit for LOB volume data over $70\%$ of trading days across all assets on the bid and ask side. 

\begin{table}[h]\footnotesize
\begin{center}
\begin{minipage}{115mm}
\tbl{Superior method fitted to volume data for six assets: 5YTN, BOBL, SP500, NIKKEI, GOLD and SILVER. MLE; Method of moments; McCullochs quantile based estimation \cite{McCulloch1986}; Pickands' estimator \cite{pickands1975statistical}; Empirical percentile method}
		{\begin{tabular}{p{4.5cm}p{2.4cm}p{1.4cm}p{1.4cm}p{1.4cm}p{1.4cm}}
		&&& \textbf{Method} \\
		\hline
		\textbf{Distribution}  & MLE  & McCullochs & Mixed L-moments & Pickands' & Empirical percentile method \\
		\hline
		1. $\alpha$-stable &       & best method & & & \\
		2. Generalized extreme value  &  best method   & &  & worst method & \\
		3. Generalized Pareto distribution  &  \textbf{*overall best method*}   &   & & second best  & failure to work  
		\end{tabular}}
		\label{tab:concdist}
\end{minipage}
\end{center}
\end{table}

\subsection{Implications for dynamic model features}\label{sec:modelRec}
To recap, a number of key findings have been observed in the LOB volume data across all assets on the bid and ask side. For a data-set to have sufficient data for modeling and be reflective of a higher frequency trading strategy, we require a sub-sampling rate of one minute or less. With a sampling rate of one minute or less, the data exhibits dependence, as shown in the Hurst exponent and the extremogram being an analog of the auto correlation function which considers upper tail dependence. When modeling the high sub-sampling rates, the resulting massive data sets can be computationally expensive when estimating parameters of a distribution.  
	
The LOB volume data at sub-sampling rates of one minute or less exhibits high levels of skewness, kurtosis and the intraday LOB volume profiles exhibit heavy tailed features. When capturing these core features of the volume process through three families of models, $\alpha$-stable, generalized Pareto distribution and generalized extreme value distribution, using different sub-sampling rates of one minute or less, we see consistent parameter estimation across all assets for the year. However, the parameters estimated across time inter-day vary for each asset considered.

A dynamic model would need to incorporate a number of key components to capture all of the features observed in this analysis. Computationally efficient methods of parameter estimation would be critical in the context of the large data sets and the higher frequency nature of application. All other statistical approaches for modeling LOB volume profiles in financial literature are invalidated with the findings of the high levels of skewness, kurtosis and the heavy tailed features in the marginal distribution. Dynamic parameter estimates would be required to capture the time varying inter day parameters. The long memory and serial dependence observed in the upper tails would need to be captured via methods such as copulas.

\section{Case study: Implications of heavy tails on liquidity (Model Based XLM)}\label{sec:casestudy}
In this section we consider an important question: 
\begin{center} \textit{What is the impact of heavy tailed features on liquidity in the LOB }? \end{center} 

The concept of liquidity aims to capture the ability to convert shares into cash, and vice-verse, at the lowest transaction costs. In \citet{harris1990liquidity}, a perfectly liquid market is defined to be one in which any amount of a given security can be instantaneously converted to cash and back to securities at no cost. In practice, this is unrealistic and a more realistic definition would consider a liquid market where the transaction costs associated with the conversion are minimized. This fundamental definition is reliant directly on the volume profiles.

There are many reasons we may want to understand the notion of LOB liquidity. From a high frequency trading perspective, firms are able to execute electronic transactions within very small time frames and take advantage of small price discrepancies. The very short time-frames used to enter and liquidate positions means that incorrect distributional assumption underlying their forecasting models on available liquidity in the LOB can very quickly cannibalize any profits made for that trade. Moving the price by under-estimating the liquidity or failing to execute enough volume when there is a larger liquidity event will degrade the quality and profitability of the strategy. Thus, key to the performance of any high frequency trading strategy are the underlying distributional assumptions made about the volumes on the LOB. 

There are regulatory frameworks in place, for example MiFID (Markets in Financial Instruments Directive) which has a key aspect of \textit{Best Execution}. The directive requires firms to take all reasonable steps to obtain the best possible result in the execution of a client order, thus appropriately considering available liquidity. When defining and measuring liquidity, firms must not only consider just the inside spread, but also market impact and opportunity costs of trading. To achieve this, the volume of resting orders in the LOB must be taken into consideration within the liquidity measure. For example, \citet{irvine2000liquidity} investigate properties of a particular class of liquidity measures, known generically as the \textit{cost of round trip trade}. \textit{Cost of round trip trade} measures summarize the structure of the LOB instantaneously for a given order size through a process of aggregation of the key features of the LOB. By construction, they are intended to capture the ex-ante committed liquidity immediately available in the market. 

Liquidity is also a feature of orderly markets, and therefore regulators have a natural interest in the dynamics of liquidity. In \cite{schiereck1995internationale} it is argued that liquidity is the most important decision-making criterion for investors in selecting the markets and assets they would like to invest in, and that it is a central concept that quantifies the quality of particular securities markets. We will demonstrate that the correct modeling of the volume profile stochastic features can have a pronounced effect on the ability to estimate and understand liquidity dynamics in the LOB. For further discussions on LOB liquidity and micro and macro-economic factors, see \citet{gomber2002market} and \citet{chordia2001market}

The Xetra Liquidity Measure (XLM), proposed by Deutsche Börse AG \citet{gomber2002market}, falls under the umbrella of \textit{cost of round trip trade} measures. Empirical studies using the XLM have compared liquidity costs across assets \citep{stange2009market} and also attempted to define and quantify liquidity risk \citep{ernst2009measuring, gomber2004zooming}. In this analysis we define a notion of \textit{Model Based XLM} which we develop to illustrate the impact of the appropriate choice of a heavy tailed volume profile model versus a unsuitable lighter tailed volume profile model currently discussed in the literature. Previous studies \citep{Biais1995,Challet2001,Maslov2001,Bouchaud2002, Gu2008, Chakrabort2010, Gould2012} on LOB volume data, underpinning the XLM measure, have primarly focused on the gamma distribution. As we have discovered in this research, the gamma distribution fails to capture the skewness and kurtosis features of the LOB volume data. Through our research we have found that the intra-day volumes exhibit heavy tails. 

In this case study, we consider the impact of marginal distributional assumptions on XLM. XLM, as described in \citet{Rosch2012market} is a volume-weighted spread liquidity measure derived from the LOB, which measures the order-size-dependent liquidity costs of a round-trip trade. To define the XLM we consider the following notation:
\begin{itemize}
\item $a$ denotes the ask, $b$ denotes the bid,

\item $P_{t}^{b,i}\in\mathbb{N}^{+}$ denotes the random variable for the limit price of the $i^{th}$ level bid at time $t$ in tick units,

\item $P_{t}^{a,i}\in\mathbb{N}^{+}$ denotes the random variable for the limit price of the $i^{th}$ level ask at time $t$ in tick units,

\item $\bm{V}_{t}^{b,i}\in\mathbb{N}^{n}$ denotes a column random vector of orders at the $i^{th}$ level bid at time $t$ . 
\end{itemize}
Using this notation we define the $p_t^m=\frac{(p_{t}^{a,1}+p_{t}^{b,1})}{2}$ the quoted midpoint, or mid price. In addition we denote the total volume available at for example the $i^{th}$ bid level by $TV_{t}^{b,i}=\bm{1}_n^T\cdot \bm{V}_{t}^{b,i}$ and $TV_{t}^{a,i}=\bm{1}_n^T\cdot \bm{V}_{t}^{a,i}$ for the ask, where $\bm{1}_n$ is a column vector of 1s. We then quantify the information present on an incoming buy limit order according to the information relating to time, price and size and an order id.: $\bm{l}^b=(l_t,l_p,l_s,\mathrm{id})$. Then the XLM at time $t$, for a certain amount of money (say US cents) denoted by R, is given by:
{\small{
\begin{equation}
\begin{split}
XLM_t(R) &=\frac{\sum _{i=1}^{k}TV_{t}^{a,i}(P_{t}^{a,i}-P_t^m)+(R-\sum ^{i=1}_{k}TV_{t}^{a,i})(P_{t}^{a,i+1}-P_t^m)}{R} \\
&\;\; + \frac{\sum _{i=1}^{k}TV_{t}^{b,i}(P_t^m-P_{t}^{b,i})+(R-\sum ^{i=1}_{k}TV_{t}^{b,i})(P_t^m-P_{t}^{b,i+1})}{R},
\end{split}
\end{equation}
}}
where $k=\max (n:\sum _{i=1}^{n}TV_{t}^{a,i}(P_{t}^{a,i}-P_t^m) < R)$.

To assess the impact of the marginal distributional assumptions on XLM, we firstly calculate the actual empirical XLM for a given asset for each sub-sample (10 second) time increment. We then estimate the parameters for the gamma distribution and the generalized Pareto distribution distribution for each level of volume one every trading day in 2010. For each corresponding time increment, we draw a realization from the associated volume profile distributions at each level of the bid and ask using the estimated parameters. Using this realization and the observed price, an estimated XLM using the gamma marginal distributional assumption and the generalized Pareto distribution marginal distributional assumption is made at each time increment with heavy and light tail categorization. If the volume on the bid and ask represents the largest 20\% of volume for that day, as defined by the POTs method, then we define the corresponding actual empirical XLM and estimated parametric Model Based XLM as \textit{heavy} tailed. 

The order size is re-calculated each day to represent the average of a half of the total volume on levels 1-5 of the bid and ask to ensure the estimation utilizes order book depth. To quantify the estimation performance using the assumption of the gamma model for the volume profiles versus the generalized Pareto distribution assumption, we utilize a forecast error measurement mean absolute percentage error, MAPE, defined as:
\begin{equation}
	MAPE = \frac{100\%}{n}\sum{\left|\frac{WS_t-\widehat{WS}_t}{\widehat{WS}_t} \right|},
\end{equation}
where $\hat{WS}_t$ denotes the model based predicted XLM.

\begin{figure}[h]
\begin{center}
\begin{minipage}{150mm}
{\subfigure[MLE method for 10 second sampling rate.]{
\resizebox*{7.5cm}{!}{\includegraphics{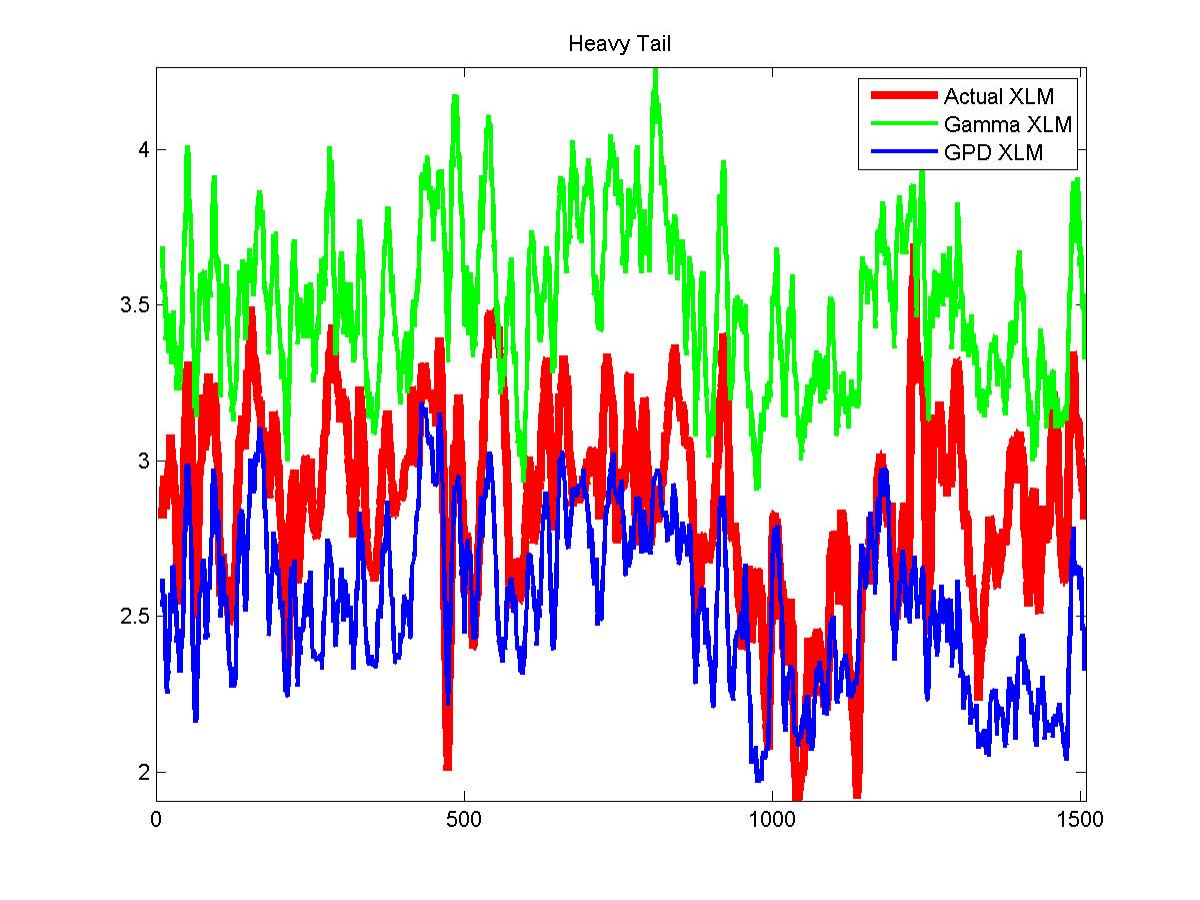}}\label{sample-figure_part_a}}
\subfigure[MLE method for 60 second sampling rate.]{
\resizebox*{7.5cm}{!}{\includegraphics{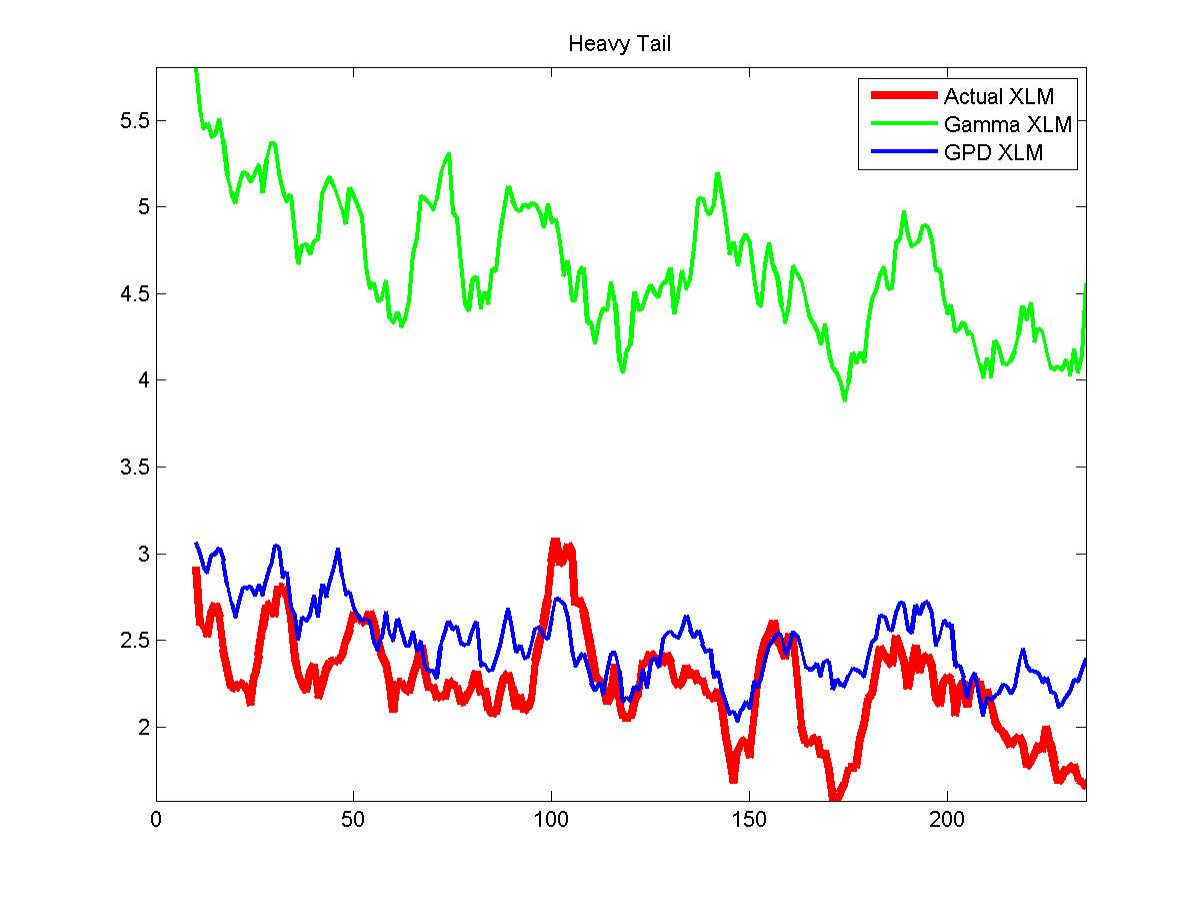}}\label{sample-figure_part_b}}
\caption{eavy tailed actual XLM, gamma estimated XLM, generalized Pareto distribution estimated XLM every 10th trading day in 2010. \textbf{Left Sub-Plots:} 10-day moving average for the 5YTN. \textbf{Right Sub-Plots:} 10-day moving average for GOLD.}
\label{fig:XLM}}
\end{minipage}
\end{center}
\end{figure}

Figures \ref{fig:XLM} left subplots show the heavy tailed actual XLM, gamma estimated XLM, generalized Pareto distribution estimated XLM using pricing from 5YTN and GOLD. It is clear from the charts that the generalized Pareto distribution estimated XLM best represents the actual XLM for both assets. The Gamma distribution fails to estimate the high volume liquidity events and thus requires more levels of the LOB to estimate the XLM, resulting in a higher estimated round trip cost of trading. This is further support by the calculation of MAPE being significantly lower (5TYN: $13\%$ and GOLD: 30\%) for generalized Pareto distribution versus Gamma (5TYN: $25\%$ and GOLD: 129\%). In addition to the underestimation of liquidity, we also see that the gamma distribution has particularly high results for MAPE GOLD. 

\FloatBarrier

\section{Conclusion}	
The aim of this paper was to create the building blocks for dynamic models, by considering the long memory in the dependence structure, the heavy tail features in the marginal distributions and the impact of sampling frequency. By implementing two procedures, the Hurst exponent and the more recent technique being the extremogram, this research confirmed the finding of long range dependence across all assets, with increasing dependence for shorter time increments. This forms a key considering for modeling the dependence structure within the LOB volume profiles. We further explored the importance of a study on heavy tailed marginal distributions in the context of the observed dependence. The large sample sizes on which to fit the extreme value distributions are sufficient to obtain consistent estimators of the marginal distribution and the heavy tailed features. 

Researching the second objective, we considered 3 distributions across 6 futures assets for an entire year of trading (2010). From this analysis we attempt to model and assess the heavy tail volume profile features present in the data at each level of the bid and ask in the LOB. In addition to the different distributional fits considered, we explored many different parameter estimation methods within each distribution to ensure that we have a method that is robust to the ultra-high frequency data analyzed and a method that best captures the features present in the data.

There was statistical evidence for the presence of sub-exponential volume profile right tails at different levels of the order book for all models considered: the $\alpha$-stable; generalized extreme value; generalized Pareto distribution models. All models demonstrated consistent results for the presence of intra-day periods of heavy tails in which the tail index parameter spiked in an asymmetric manner on level 1 of the bid and ask volume profiles. With regard to the notion that perhaps the heavy tailed features may be exchange specific due to possible effects related to specific exchange mechanisms or market participants in such exchanges, we found that the potential for heavy tailed features in the intra-day volume profiles was present in all exchanges considered in the analysis (CBOT, EUREX, SGX, CME COMEX). In addition, the asymmetry present in the times of occurrence of the heavy tailed features of the bid and ask volume profiles was also not exchange specific. Interestingly, it was certainly the case that particular asset classes had a greater tendency to systematically display heavy tailed volume profiles at all levels on the bid and ask. For instance we demonstrated that clearly the precious metals (GOLD and SILVER) contracts displayed heavy tailed intra-day behavior systematically throughout the trading year, whereas the 5YTN and BOBL displayed such features intermittently throughout the year. In addition, we took these statistical features and illustrated how they have a direct impact on forecasting and estimation of an influential measure of LOB liquidity given by the XLM, where we defined the notion of a model based XLM for forecasting.

Due to the range of high frequency sampling rates considered at 10 seconds, 5 seconds, 2 seconds and 1 second intra-day for each trading day of 2010, we were able to disambiguate the real stochastic behavior of the heavy tailed features of the LOB volume profile structures from the high-frequency ``micro-structure noise'', see recent work in \citet{ThesisMicroStruct}. We showed that for a given market and asset class, the presence of statistical features such as the heavy tailed attributes were persistent in the stochastic processes for the range of sampling resolutions. 

During non-heavy tailed volume profile intra-day time periods, we found the bid and ask LOB marginal dynamics to generally follow a strong common trend. However interestingly, during heavy tailed events, there was an asymmetry in the volume profiles on the bid and ask at all levels. It was also clear that there is strong statistical evidence from the model estimations undertaken to recommend that when developing parametric models for the volumes on the bid and ask, the models should incorporate both intra-day and inter-day dynamics for the model parameters, time varying volatility and heavy tailed features. This was clear since all assets demonstrate time variation for the parameters across the year with intra-day and inter-day variations present. For BOBL, SP500 and NIKKEI, the large number of times the shape parameter dropped below 1.8 was indicative of why we found the Normal and Gamma distributions to be a bad fit even after simple transformations were applied. Skewness for the SP500 is aligned with the findings for 5YTN, with a $\beta =1$ a large portion of the year. Both BOBL (figure ~\ref{fig:AStabBOBLPar}) and NIKKEI demonstrated left skewness, but less pronounced compared with 5YTN and SP500.

To assess the quality of the statistical model estimations, a variance weighted modified Kolmogorov-Smirnov test which attributes appropriate weights to the sub-exponential tails of the model was considered. The generalized Pareto distribution using the MLE parameter estimation method was the best fitting model. The model provided a good fit of between 70\%-100\% of trading days for all assets on both the bid and ask side. 

We further demonstrate the impact of incorrect distributional assumptions on liquidity and the commonly used liquidity measure, XLM. Previous research has suggested that a Gamma distribution is the correct marginal distribution, however we have found that it fails to account for the skewness and kurtosis in the data and as a result it grossly underestimates the volume in the tails. In the case of high liquidity events, the volume in the LOB will be underestimated, thus over-estimating the round trip cost of trading. In addition, for low count assets such as GOLD and SILVER, the incorrect gamma distributional assumption will lead to sizable LOB imbalance estimations that may lead to erroneous estimations of volume and XLM. Thus, volume forecasting models utilized by a high frequency trading strategy should account for heavy tails, time varying parameters and long memory present in the data. 

\section{Acknowledgements}
We thank Boronia Capital Pty. Ltd. for providing the data and in particular the expertise of Dr. Chris Mellen, Russell Grew and Dr David Oliver. Kylie-Anne Richards is also thankful for the support provided by the Boronia-QRSLab PhD scholarship that has support part of this research time.

\section{References}\label{refs}

\bibliographystyle{rQUF}
\bibliography{JabRef_PhD_FINAL}

\end{document}